\UseRawInputEncoding
\documentclass[pre,preprint,preprintnumbers,amsmath,amssymb]{revtex4}
\usepackage{amsmath,graphics,epsfig,epstopdf,fancyhdr,dcolumn,bm}
\usepackage{color}
\newcommand{\dv}{\Delta v_R}
\newcommand{\dvt}{\langle\Delta v^2(t)\rangle}
\newcommand{\DQL}{D_{\rm{QL}}}
\newcommand{\dvm}{\delta v_{\max}}
\begin {document}
\title{Transport of charged particles in discrete wave spectra : the importance of locality in phase velocity for quasilinear diffusion}
\author{Didier B\'enisti}
\email{didier.benisti@cea.fr} 
\affiliation{ CEA, DAM, DIF F-91297 Arpajon, France.}
\date{\today}
\begin{abstract}
In this paper, we address the motion of charged particles subjected to a discrete spectrum of electrostatic waves. We focus on situations when transport dominates, leading to significant variations in particle velocity. Nonetheless, these velocity changes remain finite due to the presence of KAM tori bounding phase space.  We analyze the conditions under which transport can be modeled as a diffusion process and evaluate the relevance of the so-called quasilinear value of the diffusion coefficient. We distinguish between traditional quasilinear diffusion, when wave-particle interaction is perturbative, and the so-called chaotic regime of diffusion, when the particle motion looks erratic.  In the perturbative regime, we demonstrate both numerically and theoretically that diffusion occurs only when wave-particle interaction is local in phase velocity; that is, when wave contributions from phase velocities far from the particles instantaneous velocities are negligible. Conversely, numerical results indicate that chaotic diffusion can occur even when wave-particle interaction is not local. However, without locality, the diffusion coefficient is not the quasilinear one. Furthermore, in regimes when quasilinear diffusion is applicable, we introduce a simple analytical expression for the time evolution of the velocity distribution function, that accounts for phase space boundaries.

\end{abstract}
\maketitle
\section{Introduction}
\label{intro}
Whether transport can be modeled by a Fokker-Planck equation remains one of the most fundamental issues in both basic and applied physics. From a foundational perspective, this amounts to addressing determinism. Indeed, the Fokker-Planck equation is mathematically equivalent to a stochastic Langevin equation. Hence  if, as shown in Refs.~\cite{benisti1,escande,elskens,elskens2}, some deterministic systems may be modeled by a Fokker-Planck equation for an arbitrary long time and with an arbitrary accuracy, there is no way for an observer to know whether such systems are indeed deterministic or stochastic. In other words, it would be utterly impossible to ascertain whether all natural systems are fully deterministic or if some irreducible degree of randomness exists. 

Furthermore, deriving the particle distribution resulting from chaotic transport is essential in countless applications. This paper focuses on transport driven by a fixed spectrum of electrostatic waves, a scenario commonly encountered in plasma physics.  The best known example is certainly the saturation regime of the warm beam-plasma instability (see for example Refs.~\onlinecite{besse} and references therein). To cite a few other examples, in space plasmas, interactions with broad wave spectra have been invoked to explain energetic ion observations in the magnetosphere \cite{ram} and electron losses in radiation belts \cite{mourenas}. Moreover, the present work is motivated by laser-plasma interaction when electrostatic waves result from stimulated Raman scattering or two-plasmon decay~\cite{kruer}. Such waves may generate energetic electrons~\cite{winjum} which are detrimental to inertial confinement fusion~\cite{icf}, or beneficial for producing X-ray sources \cite{xray} with enhanced energies.  For the application to laser-plasma interaction (and for many other ones),  the wave spectrum may certainly not be assumed stationary. However, as shown in Section~\ref{theory}, considering a fixed spectrum is enough to derive transport properties, provided that the modes growth rates vary slowly enough and that the maximum growth rate is much less than the total frequency extent of the spectrum. 

In plasma physics, transport induced by electrostatic waves is most often modeled as a quasilinear diffusion~\cite{QL}. Although quasilinear theory has been the subject of considerable debate~\cite{laval}, several papers provided a detailed discussion of its relevance for a nearly stationary spectrum (see for example Refs.~\cite{benisti1,besse,escande2,livre}). In particular, Ref.~\onlinecite{benisti1} shows that two types of diffusion may exist. In the first one, which is quasilinear and closer to the situation considered in the initial theory~\cite{QL}, the particle motion may be described by making use of a first order perturbation analysis. It may be followed by a second, so-called ``chaotic'' diffusion regime, when the particle motion looks erratic. In this second regime, the diffusion coefficient may be different from the quasilinear one, although usually of the same order. Moreover, as shown in Refs.~\cite{benisti1,elskens}, for large enough wave amplitudes, the initial quasilinear regime persists until the onset of the second, chaotic regime. In this situation, quasilinear diffusion is valid until the particles orbits reach the phase space boundaries. Now, all the aforementioned results were derived for the dynamics defined by the following Hamiltonian, assumed to be paradigmatic,
\begin{equation}
\label{Ht}
H_p=\frac{v^2}{2}+A\sum_{n=1}^N\cos(x-nt+\varphi_n),
\end{equation}
where the $\varphi_n$s were random initial phases. The key point of the derivation was the proof of locality in phase velocity for the wave-particle interaction~\cite{jsp}. Here, locality means that, at all times, transport is essentially due to those waves whose phase velocity differ by no more than $\dv\approx 6A^{2/3}$ from the instantaneous particle speed. Note that the quasilinear diffusion coefficient, as derived in Ref.~\onlinecite{QL}~for a continuous spectrum, accounts only for waves with phase velocity exactly matching the particle speed, thus representing an infinitely local interaction. This would not make sense for a discrete spectrum  and, for such a wave spectrum, the expression of the quasilinear diffusion coefficient  was derived in Ref.~\onlinecite{doxas}~and recalled in Section~\ref{theory}, Eq.~(\ref{DQL}). 

Clearly, the conclusions drawn from  the Hamiltonian $H_p$ raise several important questions, mostly (but not only) regarding their universality :
\begin{enumerate}
\item Does there exist a broad class of Hamiltonian dynamics whose transport properties are very similar to those derived from $H_p$ ?
\item Is locality a universal property in wave-particle interaction ? 
\item Does the first perturbative regime of quasilinear diffusion exist for all wave spectra ? 
\item Is there a link between locality and the first quasilinear diffusion regime ? 
\item Can chaotic diffusion exist without locality ? 
\item Talking about diffusion is somewhat abusive for a finite discrete wave spectrum because particles orbits only have access to a limited region of phase space, which raises the following question : can one derive an analytical expression for the time evolution of the velocity distribution function that accounts for phase space boundaries ?
\end{enumerate}
To the best of the author's knowledge, no previous article specifically addressed the first, fourth, and sixth issues. In Ref.~\onlinecite{besse}, Vlasov simulations showed the relevance of quasilinear theory to model the warm beam-plasma instability in the strongly nonlinear regime, while non-quasilinear evolution was observed for intermediate nonlinear regimes, thus supporting some of the results derived from the Hamiltonian $H_p$. Moreover, on a more theoretical perspective, several papers, e.g. Ref.~\onlinecite{escande2}, addressed the relevance of the initial quasilinear diffusion regime for more general wave spectra than that considered in Hamiltonian $H_p$. However, the authors of these papers implicitly made hypotheses that led to the validity of the initial quasilinear regime. In this article, we aim to address the 6 aforementioned issues for the general class of wave spectra leading to the dynamics defined by the following Hamiltonian,
\begin{equation}
\label{H}
H=\frac{v^2}{2}+A\sum_{n=1}^Na_n\cos(k_nx-\omega_nt+\varphi_n),
\end{equation}
where the $\varphi_n$s remain random initial phases, while the $a_n$s, $k_n$s and $\omega_n$s may be chosen arbitrarily. As for the pre-factor, $A$, it is  introduced as a normalization constant to allow direct comparison with the results derived from Hamiltonian $H_p$. 
For the general Hamiltonian $H$, we theoretically establish the conditions on the relative variations of the wave amplitudes, wave numbers, and wave frequencies required for locality to hold. We further show that the presence of an initial quasilinear diffusion regime is closely linked to the validity of locality. To the best of the author's knowledge, the latter issue has been completely overlooked in previous studies. 

In order to check numerically our theoretical predictions more easily, using test particle simulations, we derive the criteria that should lead to a uniform transport in phase space (at least far from the boundaries due to the presence of KAM tori). Under these criteria, if a Fokker-Planck equation is valid, it should reduce to a simple diffusion equation, $\partial_tf=D\partial^2_vf$, where $D$ is independent of $v$. In particular, we find the conditions under which the quasilinear diffusion coefficient, given by Eq.~(\ref{DQL}), is independent of $v$ and assumes the same value as for Hamiltonian $H_p$, namely $\DQL=\pi A^2/2$.  Then, quasilinear diffusion leads to very simple transport properties since $\langle\Delta v^2(t)\rangle=2 \DQL t=\pi A^2t$, where $\Delta v=v(t)-v(0)$ and where the brackets stand for the averaging over the initial phases, $\varphi_n$.  

Moerover, using numerical simulations, we test whether a chaotic diffusion regime eventually settles, regardless of the validity of locality and, when locality fails, the impact of the number of modes on the diffusion coefficient. 

Now, diffusion cannot last for all times because the particles orbits eventually hit the boundaries of phase space, as illustrated in Fig.~\ref{p2}. These boundaries are the so-called KAM tori~\cite{KAM} which exist for $v\agt\max(\omega_n/k_n)$ and $v\alt\min(\omega_n/k_n)$ and which cannot be crossed by any orbit. Consequently, one cannot simply derive the velocity distribution function by solving the diffusion equation, $\partial_tf=D\partial^2_vf$. In this paper, we provide a heuristic expression for the time variation of the distribution function, which accounts for phase space boundaries, and whose accuracy is tested against results from test particle simulations. 

The paper is organized as follows. Section~\ref{poincare} clearly defines what we mean by a ``bounded chaotic transport'', which we investigate in this article. Section~\ref{theory} presents our theoretical analysis providing the conditions  for the validity of locality and the occurence of the perturbative regime of quasilinear diffusion. It also briefly discusses the occurence of the chaotic regime of diffusion and provides conditions under which the quasilinear diffusion coefficient is uniform in phase space, thereby allowing the Fokker-Planck equation to simplify into a diffusion equation. Section~\ref{numerics} checks our theoretical predictions by making use of test particle simulations, with a particular emphasis on the link between locality and the occurence of the first quasilinear regime of diffusion. In the chaotic regime, the dependence of the diffusion coefficient on the number, $N$, of modes is further investigated.  Section~\ref{transport} provides an effective expression for the velocity distribution function resulting from a diffusion-like process limited in time by the phase space boundaries. Section~\ref{conclusion} summarizes and concludes our work. 

\section{Bounded chaotic transport}
\label{poincare}
\begin{figure}[!h]
\centerline{\includegraphics[width=12cm]{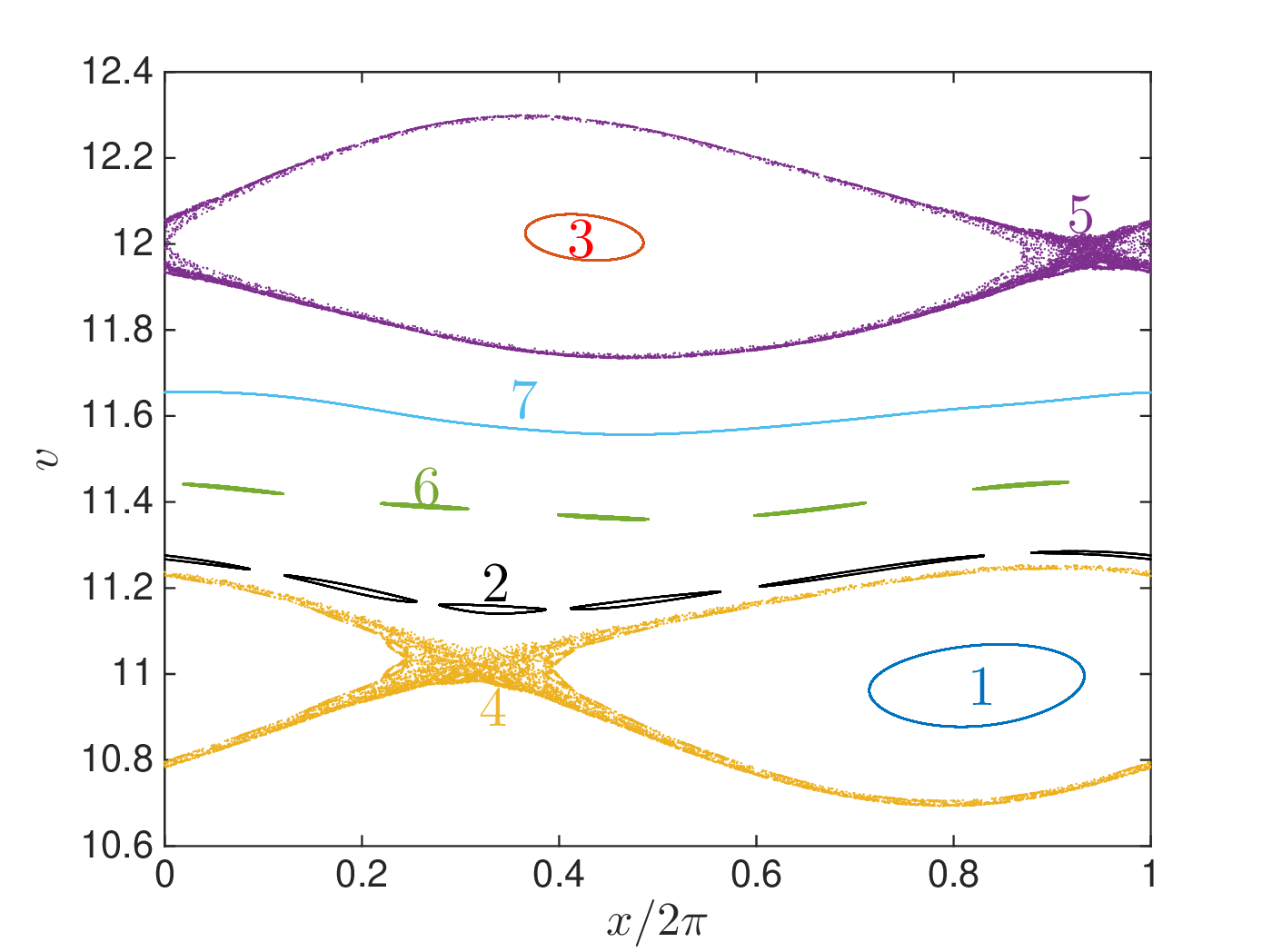}}
\caption{\label{p1}Poincar\'e sections for the Hamiltonian $H_p$ when $N=21$ and $s=0.53$. Each color corresponds to one of the initial condition 1-7, as defined by Eqs.~(\ref{pp1})-(\ref{pp7}).  In the vicinity of each Poincar\'e section, and using the same color, is indicated the corresponding initial condition. }
\end{figure}

\begin{figure}[!h]
\centerline{\includegraphics[width=12cm]{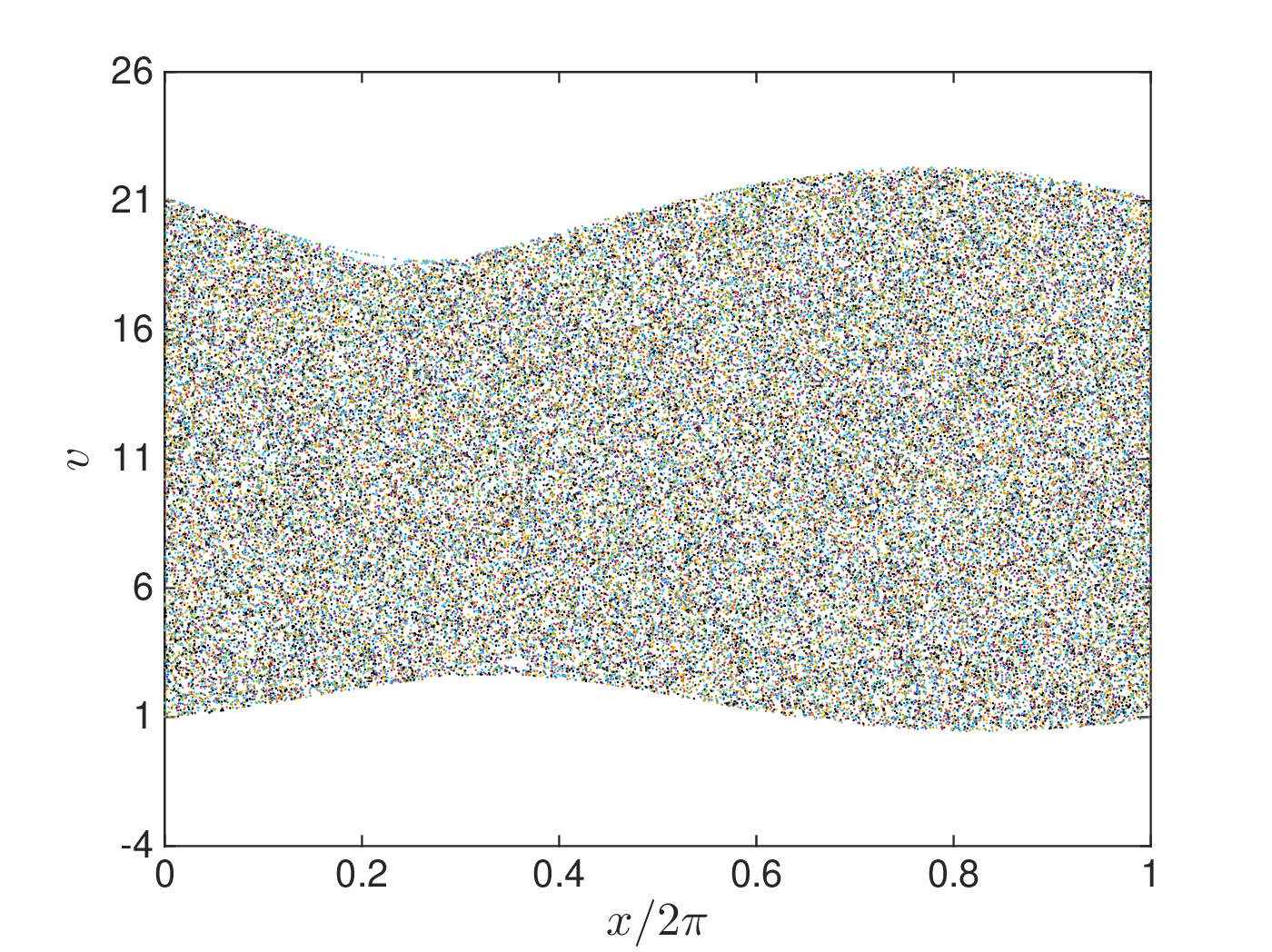}}
\caption{\label{p2}Same as Fig.~\ref{p1} for $s=4$.}
\end{figure}

In this section, we qualitatively  discuss the physical effect we aim to quantify : a bounded chaotic transport in velocity. The discussion is based on Poincar\'e sections for the dynamics ruled by the Hamiltonian $H_p$, Eq.~(\ref{Ht}). This amounts to plotting the phase space coordinates $\left[x\left(t_n\right),v\left(t_n\right)\right]$ at discrete times $t_n=2\pi n$. The positions $x(t_n)$ are plotted modulo $2\pi$ so that $0\leq x(t_n)\leq 2\pi$. {In Figs.~\ref{p1}~and~\ref{p2}, the phase space coordinates are computed by solving numerically the Hamilton equations for $H_p$ using a leapfrog integrator~\cite{verlet} with the time step $\delta t=2\pi\times10^{-3}$, while the total integration time is $\Delta t=2\pi\times10^{4}$. }

\begin{figure}[!h]
\centerline{\includegraphics[width=12cm]{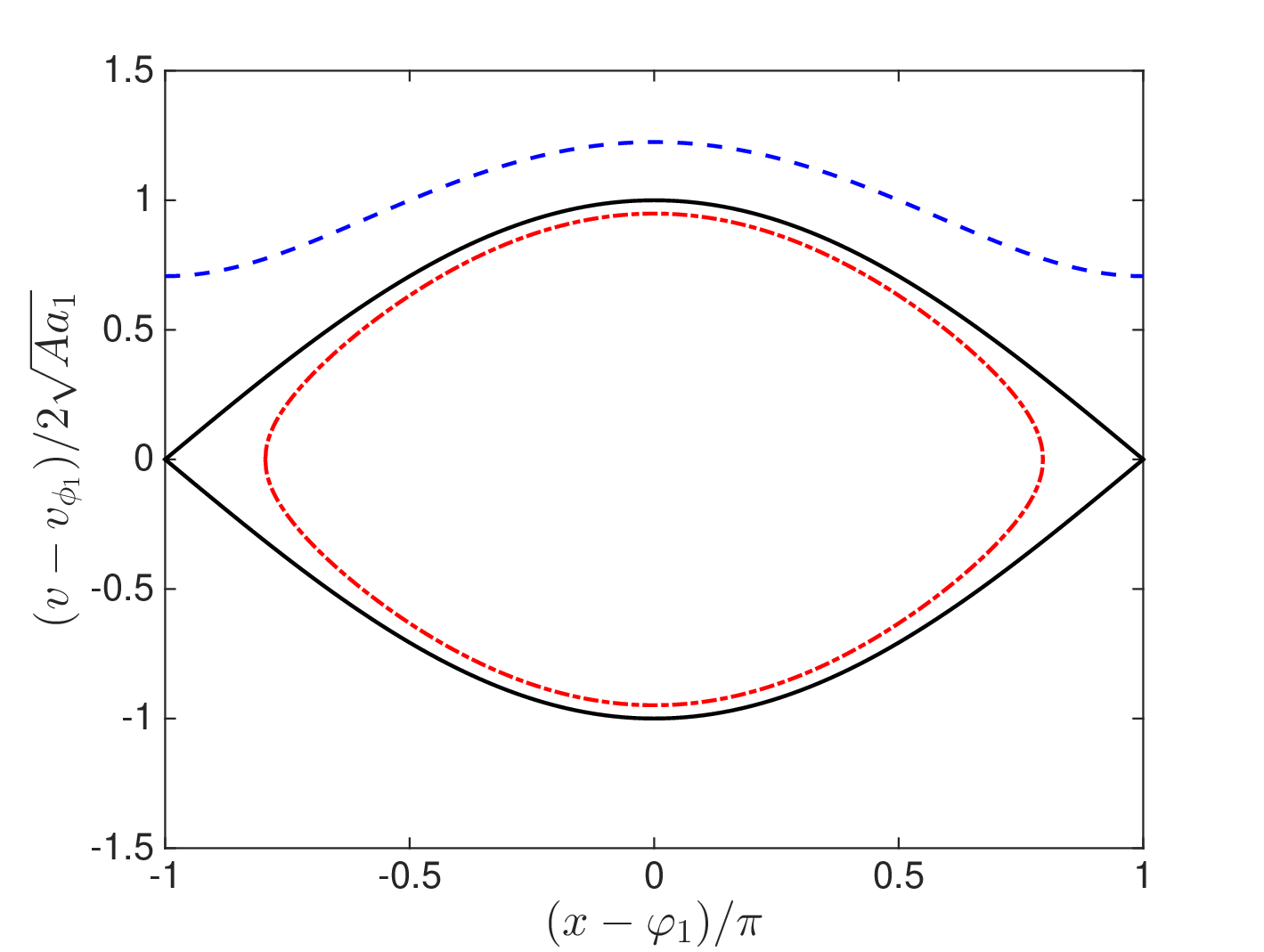}}
\caption{\label{pendule}Phase portrait of the pendulum. The blue dashed line is an untrapped orbit, the red dashed-dotted line is a trapped orbit and the black solid line is the separatrix. }
\end{figure}

When $N=1$, the dynamical system is equivalent to a pendulum. The corresponding phase portrait, shown in Fig.~\ref{pendule}, contains two types of orbits : trapped and untrapped ones, separated by a curve called the separatrix. This separatrix is centered at $v = v_{\phi_1}$ and its width is $4\sqrt{Aa_1}$. 

When $N>1$, if the phase portrait were a mere superposition of $N$ pendulum-like phase portraits, it would consist of $N$ sets of trapped and untrapped orbits, separated by $N$ separatrices, centered about $v=v_{\phi_n}$ and of width $4\sqrt{Aa_n}$. Clearly, this picture has no chance to be correct when the separatrices overlap. This happens when the Chirikov overlap parameter,
 \begin{equation}
\label{chirikov}
s(v)=2\sqrt{A}Ê\frac{\sqrt{a_{n^*}}+\sqrt{a_{n^*+1}}}{\delta v_{\phi_{n^*}}},
\end{equation}
is larger than unity. In Eq.~(\ref{chirikov}), the indices are ordered in such a way that $v_{\phi_n}\equiv \omega_n/k_n$ is an increasing function of $n$ and  $\delta v_{\phi_{n^*}}=v_{\phi_{n^{*}+1}}-v_{\phi_{n^{*}}}$, where $n^{*}$ is such that $v_{\phi_{n^*}}\leq v\leq v_{\phi_{n^*+1}}$. Note that for $H_p$ and when $\min(v_{\phi_n})+2\sqrt{A}< v<\max(v_{\phi_n})-2\sqrt{A}$, $s=4\sqrt{A}$ is uniform in $v$.

The qualitative change in the phase portrait as $s$ increases is illustrated in Figs.~\ref{p1} and \ref{p2}, corresponding to $s = 0.53$ and $s = 4$, respectively. These figures plot 7 different Poincar\'e sections of the Hamiltonian $H_p$ with $N=21$, each corresponding to one of the following seven initial conditions,
\begin{eqnarray}
\label{pp1}
1 : && (x_1=\pi-\varphi_{11} ; v_1=11+\sqrt{A}/2), \\
2 : &&(x_2=\pi-\varphi_{11} ; v_2=11+2.05\sqrt{A}), \\
3 : &&(x_3=\varphi_{12} ; v_3=12+\sqrt{A}/2), \\
4 : &&(x_4=-\varphi_{11} ; v_4=11), \\
5 : &&(x_5=\varphi_{12} ; v_5=12), \\
6 : &&(x_6=\pi\varphi_{11} ; v_6=11.43), \\
\label{pp7}
7 : &&(x_7=\varphi_{11} ; v_7=11.62).
\end{eqnarray}


The first and third initial conditions lie, respectively, within the separatrices associated with $n=11$ and $n=12$, so that the corresponding orbits are expected to be the counterparts of trapped orbits for the pendulum.  The second initial condition is located above the separatrix for $n=11$,  so that the corresponding orbit is expected to be the counterpart of an untrapped orbit for the pendulum. The fourth and fifth initial conditions are situated, respectively, at the X-points of the separatrices corresponding to $n=11$ and $n=12$. As for the sixth and seventh initial conditions, depending on the value of $A$, they may lie either within or above the separatrix corresponding to $n=11$. 

In Fig.~\ref{p1}, when $s=0.53$, there are indeed pendulum-like orbits. Trapped ones with initial conditions 1, 3, and an untrapped one with initial condition 7. These are so-called KAM tori~\cite{KAM}, i.e., orbits slightly deformed compared to those of the pendulum. In particular, the  orbit from initial condition 7 spans the whole $[0,2\pi]$ interval and cannot be crossed by any other orbit because of causality. As a result, any orbit starting in a region bounded by two such KAM tori remains permanently confined in this region, which strongly reduces transport.   Other initial conditions lead to more complicated orbits. The initial conditions 2 and 6 do not lead to pendulum-like orbits but to chains of islands. As for the initial conditions 1 and 5, they do not lead to one-dimensional orbits but to so-called stochastic layers, located about deformed pendulum-like separatrices. The existence of stochastic layers is the illustration of some, very limited, chaotic transport. Indeed, when $s=0.53$, the change in velocity remains very small for all the considered initial conditions. 

The situation changes dramatically when $s=4$, as illustrated in Fig.~\ref{p2}. Here, the phase portrait looks like one big stochastic layer. Indeed, for all the considered initial conditions, the orbits seem to densely fill a region of phase space located between $v\approx\min(v_{\phi_n})=1$ and $v\approx\max(v_{\phi_n})=21$. Hence, there is transport since velocities change significantly for all the considered initial conditions. However, this transport is limited because an orbit can only access a region of phase space bounded by KAM tori similar to that plotted in Fig.~\ref{p1}~for the initiation condition 7. This is what we call a bounded chaotic transport and, in the remainder of this paper, we only consider situations similar to that illustrated in Fig.~\ref{p2}. 
\section{Theoretical analysis}
\label{theory}
\subsection{Validity of locality}
\label{locality}
In this Paragraph, we discuss the relevance of locality for the Hamiltonian $H$ defined by Eq.~(\ref{H}). To do so, we use the same procedure as in Ref.~\onlinecite{jsp}, i.e., we make use of a perturbation analysis about a given velocity, $v_0$. This lets us conclude whether there exists a range in velocity, $\dv$, such that the waves with phase velocity $v_\phi$ satisfying $\vert v_0-v_\phi\vert>\dv$ have, on the average, a negligible impact on transport. Before proceeding to the analysis, we normalize velocities to the, still unknown, range in velocity, $\dv$. Hence, we define the following new variables,
\begin{eqnarray}
v_1&=&v/\dv,\\
x_1&=&x,\\
\tau&=&\dv t.
\end{eqnarray}
In these variables the dynamics derives from the Hamiltonian,
\begin{equation}
\label{H1}
H_1=\frac{v_1^2}{2}+\frac{A}{\dv^2}\sum_na_n\cos(k_nx_1-\omega_n\tau/\dv+\varphi_n).
\end{equation}
Now, to prove locality, we need to prove that there exists a canonical change of variables $(x_1,v_1)\to(x',v')$, defined from a generating function, $\Phi(x_1,v',t)\equiv x_1v'+\sum_l \varepsilon^l\Phi_l$ (the small parameter of the expansion, $\varepsilon$, being still to define) such that, in the new variables, 
\begin{eqnarray}
\label{xp}
x'&=&x_1+\partial_{v'}\Phi\\
\label{vp}
v'&=&v_1-\partial_{x_1}\Phi,
\end{eqnarray}
the new Hamiltonian
\begin{equation}
\label{Hp}
H'=H_1+\partial_\tau\Phi,
\end{equation}
reads
\begin{equation}
\label{H2p}
H'=H'_0(v')+\sum_{\vert\varpi_n/\kappa_n-v_0\vert\leq\dv}\mathcal{A}_n\cos\left(\kappa_nx'-\varpi_nt+\psi_n\right)+\varepsilon^{l+1}R_{l+1},
\end{equation}
where the $\mathcal{A}_n$s, $\kappa_n$s, $\varpi_n$s and $\psi_n$s are amplitudes, wave numbers, frequencies and phases which are derived from the perturbative expansion, and where there exists an upper bound for the remainder, $R_l$, independent of the number of modes, $N$. For the Hamiltonian $H_p$, Eq.~(\ref{Ht}), considered in Ref.~\onlinecite{jsp}, the terms of the perturbative expansion could be bounded by those of a Gevrey series, which allowed to provide an exponentially small estimate for $\varepsilon^{l+1}R_{l+1}$, with $\varepsilon=A/\dv^{3/2}$. 

Using Eqs.~(\ref{xp})-(\ref{Hp}), one finds, at first order,
\begin{equation}
\label{H3p}
H'=\frac{v^{'2}}{2}+v'\frac{\partial\Phi_1}{\partial x_1}+\frac{A}{\dv^2}\sum_na_n\cos(k_nx_1-\omega_n\tau/\dv+\varphi_n)+\frac{\partial \Phi_1}{\partial t}+\frac{1}{2}\left(\frac{\partial \Phi_1}{\partial x_1}\right)^2,
\end{equation}
and we want $\Phi_1$ to be solution of the following equation,
\begin{equation}
\label{1}
v'\frac{\partial\Phi_1}{\partial x_1}+\frac{\partial \Phi_1}{\partial t}=-\frac{A}{\dv^2}\sum_{\vert\omega_n/k_n-v_0\vert>\dv}a_n\cos(k_nx_1-\omega_n\tau/\dv+\varphi_n).
\end{equation}
Then, if the remainder $(1/2)\left(\partial_{x_1}\Phi_1\right)^2$ is negligible, the Hamiltonian $H'$ essentially accounts for the action of the waves whose phase velocity, $v_{\phi_n}=\omega_n/k_n$, is such that $\vert v_{\phi_n}-v_0\vert\leq\dv$. The solution of Eq.~(\ref{1}) is easily found to be,
\begin{equation}
\label{phi1}
\Phi_1=-\frac{A}{\dv^2}\sum_{\vert\omega_n/k_n-v_0\vert>\dv}\frac{a_n\sin(k_nx_1-\omega_n\tau/\dv+\varphi_n)}{k_nv'-\omega_n/\dv}.
\end{equation}
Then, at first order, the remainder is
\begin{equation}
\frac{1}{2}\left(\frac{\partial \Phi_1}{\partial x_1}\right)^2=\frac{A^2}{4\dv^4}\sum_{n_1}\sum_{n_2}\frac{a_{n_1}a_{n_2}\left[C_++C_-\right]}{\Delta_{n_1}\Delta_{n_2}},
\end{equation}
where
\begin{eqnarray*}
C_+&=&\cos[(k_{n_1}+k_{n_2})x_1-(\omega_{n_1}+\omega_{n_2})\tau/\dv+\varphi_{n_1}+\varphi_{n_2}]\\
C_-&=&\cos[(k_{n_1}-k_{n_2})x_1-(\omega_{n_1}-\omega_{n_2})\tau/\dv+\varphi_{n_1}-\varphi_{n_2}],\\
\Delta_{n_i}&=&(v'-v_{\phi_{n_i}}/\dv).
\end{eqnarray*}
Note that $\Delta_{n_i}>1$ when $v'=v_0/\dv$. Since we are only interested in statistical properties then, just like in Ref.~\onlinecite{jsp}, we estimate the remainder by its mean value with regards to the phases $\varphi_n$. When $v'\approx v_0/\dv$, 
\begin{equation}
\label{eq15}
\left\langle \frac{1}{2}\left(\frac{\partial \Phi_1}{\partial x_1}\right)^2\right\rangle\approx\frac{A^2}{4\dv^2}\sum_{n}\frac{a_n^2}{(v_0-v_{\phi_n})^2}.
\end{equation}
Let us now choose a velocity scale, $\delta v$, and let us introduce,
\begin{equation}
\label{II-7}
A_l^2=\frac{A^2}{\delta v}\sum_{i_l}a_{i_l}^2,
\end{equation}
where the sum is over the terms such that
\begin{equation}
\label{n}
\dv+\frac{(2l-1)\delta v}{2}\leq\vert v_{\phi_{i_l}}-v_0 \vert\leq\dv+\frac{(2l+1)\delta v}{2}.
\end{equation}
When $\delta v$ is small enough,
\begin{eqnarray}
\nonumber
\left\langle\frac{1}{2}\left(\frac{\partial \Phi}{\partial x_1}\right)^2\right\rangle&\approx& \frac{1}{4\dv^3}\frac{\dv}{\delta v}\sum_{l}\frac{A_l^2}{(\dv/\delta v+l)^2}\\
\label{II-8}
&\approx& \frac{1}{4\dv^3}\frac{\dv}{\delta v}\int_{x_{\min}}^{x_{\max}}\frac{A^2(x)}{(\dv/\delta v+x)^2}dx,
\end{eqnarray}
where $x_{\min}=\min(l)$ and $x_{\max}=\max(l)$ for each $l$ fulfilling Eq.~(\ref{n}). Clearly, the property of locality is satisfied if one can find an upper bound for the remainder which is independent of the extent, in phase velocity, of the wave spectrum. This is true if the integral in the right hand side of Eq.~(\ref{II-8}) converges when $x_{\max}\to\infty$. {For a regular enough variation of $A^2(x)$ [for example when $A^2(x)/x$ is a monotonous function of $x$],} this happens only when $A^2(x)/x$  goes to zero when $x$ goes to infinity, i.e., $A^2(v_\phi)/v_\phi$ goes to zero when $v_\phi$ goes to infinity. Hence, the property of locality is not always true, depending on the wave spectrum. 

Now, if $A^2(v_\phi)/v_\phi$ goes to zero quickly enough and if $A^2(v_\phi)$ remains nearly constant for $v_\phi\approx v_0$, so that $A^2(v_\phi)\approx \overline{A_v^2}$, where $\overline{A^2_v}$ is the mean value of $A^2(v_\phi)$ over a velocity scale of the order of $\dv$, Eq.~(\ref{II-8}) reads, 
\begin{equation}
\label{final1}
\left\langle\frac{1}{2}\left(\frac{\partial \Phi}{\partial x_1}\right)^2\right\rangle\approx \frac{\overline{A^2_v}}{4\dv^3}.
\end{equation}
Then, for the remainder to be negligible, one needs to choose,
\begin{equation}
\label{Deltav}
{\dv=\alpha\left(\overline{A^2_v}\right)^{1/3},}
\end{equation}
{with $\alpha\gg1$,} which lets naturally choose $\varepsilon=\sqrt{\overline{A^2_v}}/\dv^{3/2}$ for the small parameter of the perturbation analysis. This is a plain generalization of the small parameter introduced in Ref.~\onlinecite{jsp}~for the Hamiltonian $H_p$.  
\subsection{Quasilinear diffusion}
\label{diffql}
\subsubsection{The quasilinear diffusion coefficient for a discrete wave spectrum}
\label{ddqql}
As is well known~\cite{benisti1}, when the velocity distribution function obeys a diffusion equation, $\partial_tf=D\partial_{v^2}f$, $D$ being independent of $v$, 
\begin{equation}
\label{diff}
\langle\Delta v^2(t)\rangle=2Dt,
\end{equation}
where $\Delta v(t)=v(t)-v(0)$. Moreover, there is quasilinear diffusion, in the spirit of the quasilinear theory as introduced in Ref.~\onlinecite{QL}, if Eq.~(\ref{diff}) is fulfilled in the regime when $x(t)\approx x(0)+v(0)t$. For the sake of simplicity (and without loss of generality), we assume that all particles have the same initial velocity, $v(0)=v_0$. Then, it is easily shown that,
\begin{equation}
\label{dv2}
\langle\Delta v^2\rangle\approx \frac{A^2}{2}\int_0^t\int_0^t \sum_{n}k_n^2a_n^2\cos[(k_nv_0-\omega_n)(t_1-t_2)]dt_1dt_2.
\end{equation}
Let us now choose a frequency scale, $\delta \omega$, and let us introduce,
\begin{equation}
\label{II-14}
B_l^2=\frac{A^2}{\delta \omega}\sum_{i_l}k_{i_l}^2a_{i_l}^2, 
\end{equation}
where the sum is over the terms such that,
\begin{equation}
\label{II-14-b}
\frac{(2l-1)\delta \omega}{2}\leq\vert k_{i_l}v_0-\omega_{i_l}\vert\leq\frac{(2l+1)\delta \omega}{2}.
\end{equation}

Then, for $\delta \omega$ small enough,
\begin{eqnarray}
\nonumber
\langle\Delta v^2\rangle&\approx& \frac{\delta \omega}{2}\int_0^t\int_0^t \sum_{l}B_l^2\cos[l\delta\omega(t_1-t_2)]dt_1dt_2,\\
\label{II-16}
&=&\frac{1}{2\delta\omega}\int_0^{\delta \omega t}\int_0^{\delta \omega t} \sum_{l}B_l^2\cos[l(t'_1-t'_2)]dt'_1dt'_2.
\end{eqnarray}
Let us first consider the situation when, for a large enough number of terms $l$ about zero, $B_l^2$ remains nearly constant, $B_l^2\approx\overline{B^2}$, and when the contribution of the other terms is negligible. For definiteness, let us introduce $L\gg 1$ such that $B_l^2\approx\overline{B^2}$ whenever $-L\leq l\leq L$. Then, when $\vert t'_1-t'_2\vert\agt 2\pi/L$, $\sum_{l}B_l^2\cos[l(t'_1-t'_2)]\approx 2\pi\overline{B^2}\delta_ {2\pi}(t'_1-t'_2)$, where $\delta_{2\pi}(t)$ is the $2\pi$-periodic Dirac distribution. In this situation, 
\begin{equation}
\label{new1}
\langle\Delta v^2\rangle\approx\pi \overline{B^2} t,
\end{equation}
whenever $2\pi/L\alt t<2\pi/\delta \omega$. 

In the opposite situation, when the spectrum is very peaked about some $B_l$s, it is quite clear from Eq.~(\ref{II-16}) that the time evolution of $\dvt$ is not diffusive. In this case, wave-particle interaction is dominated by trapping within a few modes, a scenario out of the scope of this paper and which is, henceforth, excluded. For the remainder of this article, the wave spectrum is is implicitly assumed to be smooth.

Eq.~(\ref{new1}) defines the quasilinear diffusion coefficient as
\begin{equation}
 \DQL=\frac{\pi\overline{B^2} }{2}.
\end{equation}
Note that, if $k_n$ remains nearly constant over a large enough range in $n$, $k_n\approx \overline{k}$, then $B^2\approx \overline{k}\overline{A^2}$, so that Eq.~(\ref{Deltav}) reads, 
\begin{equation}
\dv=\alpha\left( \DQL/\overline{k}\right)^{1/3},
\end{equation}
{and is reminiscent of the results previously obtained by Dupree~\cite{dupree}, although derived in a completely different sprit. Indeed, Dupree assumes diffusion in velocity in order to derive locality, which is not done here. Moreover, as discussed in Section~\ref{numerics}, there may be diffusion even when wave-particle interaction is not local. }
Now, 
\begin{equation}
B^2\sim A^2 \left\vert\frac{k_n^2a_n^2}{\delta(k_nv_0-\omega_n)}\right\vert= \left\vert\frac{k_n^2a_n^2}{v_0\delta k_n-\delta\omega_n}\right\vert.
\end{equation}
Since we only keep the terms $n$ such that $v_0\approx v_{\phi_n}=\omega_n/k_n$, 
\begin{eqnarray}
\nonumber
B^2&{\approx}&{\left\vert\frac{k_n^2A^2a_n^2}{\omega_n\delta k_n/k_n-\delta\omega_n}\right\vert }\\
\nonumber
&{=}&{\left\vert\frac{k_nA^2a_n^2}{-\omega_n\delta k_n/k_n^2+\delta\omega_n/k_n}\right\vert}\\
\label{II-22}
&{=}&{\left\vert\frac{k_nA^2a_n^2}{\delta v_{\phi_n}}\right\vert.}
\end{eqnarray}Hence,
\begin{equation}
\label{DQL}
 \DQL=\frac{\pi}{2}\left\vert \frac{k_nA^2a_n^2}{\delta v_{\phi_n}}\right\vert_{v_{\phi_n}\approx v},
\end{equation}
{which yields $\DQL=\pi A^2/2$ for the Hamiltonian dynamics defined by $H_p$. }

Coming back to physical units, if the total electrostatic field of the wave spectrum is $E=\sum_nE_n\cos(k_nx-\omega_n+\varphi_n)$, then $d_tv=(q/m)\sum_nE_n\cos(k_nx-\omega_n+\varphi_n)$ where $q$ and $m$ are the charge and mass of the considered particles. From the Hamilton equations for $H$, Eq.~(\ref{H}), $d_tv=\sum_nk_nAa_n\cos(k_nx-\omega_n+\varphi_n)$. Hence, $Aa_n=(q/m)(E_n/k_n)$ and Eq.~(\ref{DQL}) reads,
\begin{equation}
 \DQL=\frac{\pi}{2} \left(\frac{q}{m}\right)^2\left\vert\frac{E_n^2}{k_n\delta v_{\phi_n}}\right\vert_{v_{\phi_n}\approx v},
\end{equation}
which is the same formula as that already published in Ref.~\onlinecite{doxas}. 

\subsubsection{Validity of quasilinear diffusion and connection with locality}
\label{validity}
The remainder of the perturbative expansion, derived at first order and given by Eq.~(\ref{eq15}), reads 

\begin{equation}
\label{eq16}
\left\langle \frac{1}{2}\left(\frac{\partial \Phi_1}{\partial x_1}\right)^2\right\rangle\approx\frac{A^2}{4\dv^2}\sum_{n=1}^N\frac{(k_na_n)^2}{(k_nv_0-\omega_n)^2}.
\end{equation}
One can find an upper bound for this remainder independent of the number of modes, $N$, provided that the series in the right-hand side of Eq.~(\ref{eq16}) converges. When $\omega_n/k_n$ increases with $n$, this amounts to requiring that the series $\sum_n \left(k_na_n/\omega_n\right)^2$
converges, while when $\omega_n/k_n$ decreases with $n$, the series $\sum_na_n^2$
 has to converge.

Moreover, a direct integration of Eq.~(\ref{dv2}) leads to,
\begin{equation}
\label{dv2b}
\langle\Delta v^2\rangle= A^2\sum_{n=1}^N\frac{k_n^2a_n^2\left\{1-\cos\left[(k_nv_0-\omega_n)t\right]\right\}}{(k_nv_0-\omega_n)^2}.
\end{equation}
In order to derive quasilinear diffusion in Paragraph~\ref{ddqql}, we had to assume that the dominant contribution to $\langle\Delta v^2\rangle$ was from the terms $n$ such that $v_{\phi_n}\approx v_0$, and that these terms all had about the same amplitude. This may only be true if the series $\sum_n\left[k_na_n/(k_nv_0-\omega_n)\right]^2$ converges. This is exactly the same condition as Eq.~(\ref{eq16}) for locality. Consequently, the initial quasilinear regime of diffusion {(which may only occur after a time interval larger of of the order of the inverse of the total frequency spread of the wave spectrum~\cite{note})} is only expected when wave-particle interaction is local. 

 Let us now consider the situation when the modes amplitudes $a_n$ depend on time so that $a_n\equiv a_n(t)$. Let us, moreover, assume that the modes growth rates, $\Gamma_n=d_t\ln(\vert a_n\vert)$, vary slowly enough for their time derivatives to be neglected when solving Eq.~(\ref{1}) on $\Phi_1$. Then, considering time varying amplitudes simply amounts to considering complex frequencies and Eq.~(\ref{eq16}) remains valid provided that the denominator in its right-hand side be changed into $(k_nv_0-\omega_n)(k_nv_0-\omega_n^*)$. Consequently, if the maximum growth rate is much less than the frequency width of the spectrum, the conclusion regarding the validity of locality remains unchanged.  However, since $a_n$ now has to be understood as a function of time, the validity of locality may also be time dependent. 
 
 Under the same hypotheses for $\Gamma_n$ as previously, Eq.~(\ref{dv2b}) is changed into
 \begin{equation}
\label{dv2c}
\langle\Delta v^2\rangle= \frac{A^2}{2}\sum_{n=1}^N\frac{k_n^2a_n^2\left\{1-2r_n(t)\cos(k_nv_0-\omega_n)t+r_n^2(t)]\right\}}{(k_nv_0-\omega_n)(k_nv_0-\omega_n)^*},
\end{equation}
where we have denoted $r_n(t)=a_n(0)/a_n(t)$. Regardless of the value of $r_n(t)$, the condition for the validity of the initial quasilinear regime of diffusion is the same as for locality. Hence, the conclusions for a time varying spectrum are the same as for a fixed one.

\subsection{Chaotic diffusion}
In this Paragraph, we quickly discuss the so-called chaotic regime of diffusion, although an in-depth investigation of this regime is way beyond the scope of this paper. More specifically, we assume that the position, $x(t)$, may be considered as a stochastic process with finite correlation time, which we let go to zero. Then, using the same definition as in Paragraph~\ref{diffql}, we find, 
\begin{equation}
\label {II-19}
\langle\Delta v^2(t)\rangle\approx\frac{1}{2\delta\omega}\int_0^{\delta \omega t}\int_0^{\delta \omega t}\langle\cos\{k_n[x(\delta\omega t'_1)-x(\delta\omega t'_2)]\}\rangle \sum_{n}B_n^2\cos[n(t'_1-t'_2)]dt'_1dt'_2.
\end{equation}
Again if, for a large enough range in $n$ about zero, $B_n^2$ remains nearly constant, $B_n^2\approx\overline{B^2}$, and if the contribution from the other terms is negligible, $\sum_{n}B_n^2\cos[n(t'_1-t'_2)]\approx 2\pi\overline{B^2}\delta_{2\pi}(t_1-t_2)$. Moreover, if $\langle\cos\{k_n[x(\delta\omega t'_1)-x(\delta\omega t'_2)]\}\rangle \approx 0$ whenever $\vert t_1-t_2\vert>2\pi/\delta\omega$, one finds $\Delta v^2(t)\approx 2  \DQL t$ whatever $t>0$~\cite{note}. Note that the same result holds for a time-varying spectrum provided that the $B_n$s do not vary significantly over a timescale of the order of the inverse of the total frequency width. Just like in Paragraph~\ref{diffql}, under this condition, we recover the same result as for a fixed spectrum. Consequently,  in the remainder of this article and for the sake of simplicity, we only consider stationary spectra.

Now, if the aforementioned conditions are not fulfilled, it is not easy to conclude. In Ref.~\onlinecite{benisti1}, we used locality to prove chaotic diffusion. However, we did not prove that locality was compulsory for chaotic diffusion to occur. This issue will be investigated numerically in Section~\ref{numerics}.

\subsection{Conditions for a uniform transport}
The theoretical predictions of Paragraph~\ref{diffql}, together with the occurence of chaotic diffusion, are tested numerically in a more transparent fashion when chaotic transport is uniform over a large region of phase space. Indeed, in this case, provided that a Fokker-Planck equation is valid, this equation simplifies into a diffusion equation, and $\langle\Delta v^2\rangle$ is expected to evolve linearly in time. 

We assume that transport is uniform provided that the quasilinear diffusion coefficient, defined by Eq.~(\ref{DQL}), and the overlap parameter defined by Eq.~(\ref{chirikov}),
 are independent of $n$. We seek to fulfill these conditions on $ \DQL$ and $s$ for the following variations of $k_n$, $a_n$ and $\omega_n$, 
 \begin{eqnarray}
\label{2b-1}
k_n&=&k_0n^\alpha,\\
\label{2b-2}
a_n&=&a_0n^\beta,\\
\label{2b-3}
\omega_n&=&\omega_0n^\gamma.
\end{eqnarray}
 One easily finds that $ \DQL=\pi A^2/2$ and $s=4\sqrt{A}$ when,
\begin{eqnarray}
\label{2b-29}
\alpha&=&\frac{3(\gamma-1)}{2}\\
\label{2b-30}
\beta&=&1-\gamma,\\
\label{2b-31}
a_0\omega_0&=&\frac{2}{3-\gamma},\\
\label{2b-32}
k_0&=&a_0^{-3/2}.
\end{eqnarray}
When Eqs.~(\ref{2b-29})-(\ref{2b-32}) are fulfilled, the transport properties of the Hamiltonian $H$ defined by Eq.~(\ref{H}) should be very similar to those of the Hamiltonian $H_p$ defined by Eq.~(\ref{Ht}). We also want to impose a similar time evolution for the dynamics derived from these Hamiltonians. To this end, we introduce $n^*$ such that, $v(0)\approx \omega_{n^*}/k_{n^*}$. Then, making use of the expansions
\begin{eqnarray}
k_n&\approx&k_{n^*}+(dk_n/dn)_{n=n^*} (n-n^*)=k_{n^*}+\alpha k_{n^*}(n-n^*)/n^*,\\
\omega_n&\approx&\omega_{n^*}+(d\omega_n/dn)_{n=n^*}(n-n^*)=\omega_{n^*}+\gamma\omega_{n^*}(n-n^*)/n^*,\\
x(t)&\approx& x(0)+v(0)t\approx x(0)+\left(\omega_{n^*}/k_{n^*}\right)t,
\end{eqnarray}
leads to,
\begin{equation}
k_nx-\omega_nt\approx \frac{(\alpha-\gamma)\omega_0}{n^*}(n-n^*)t.
\end{equation}
In order for the time evolution to be as similar as possible to that of Hamiltonian $H_p$ defined by Eq.~(\ref{Ht}), we choose $\omega_0$ so that 
\begin{equation}
\label{cond}
\frac{(\alpha-\gamma)\omega_0}{n^*}=1.
\end{equation}
Using Eqs.~(\ref{2b-29})-(\ref{2b-32}), the condition given by Eq.~(\ref{cond}) reads, when $\gamma\neq3$,
\begin{equation}
\label{w0}
\omega_0=\left[v(0)\right]^{1-\gamma}\left(\frac{2}{3-\gamma}\right)^\gamma.
\end{equation}

Now, from Eqs.~(\ref{2b-29})-(\ref{2b-32}), $\omega_n/k_n=(\omega_0/k_0)n^{(3-\gamma)/2}$ and increases with $n$ provided that $\gamma<3$. In this case, $(k_na_n/\omega_n)^2=(k_0a_0/\omega_0)^2n^{-1-\gamma}$ so that the series $\sum_n(k_na_n/\omega_n)^2$ converges whenever $\gamma>0$. When $\gamma>3$, one has to investigate the series $\sum_na_n^2=a_0^2\sum_nn^{3(\gamma-1)}$, which diverges for such values of $\gamma$. Hence, from Eq.~(\ref{eq16}), the property of locality and the initial quasilinear regime are expected to be valid whenever 
\begin{equation}
\label{gamma}
0<\gamma<3. 
\end{equation}
Note that the situation modeled by the Hamiltonian $H_p$, defined by Eq.~(\ref{Ht}), corresponds to $\gamma=1$. For this value of $\gamma$ we proved both theoretically and numerically in Refs.~\cite{jsp,benisti1} the property of locality and the initial quasilinear regime, in agreement with the condition Eq.~(\ref{gamma}). 
\section{Numerical results}
\label{numerics}
In this section, we make use of numerical simulations to test the theoretical predictions of Section~\ref{theory}, i.e., that the property of locality and the initial quasilinear regime are valid only when the series in the right-hand of Eq.~(\ref{eq16}) converges when $N\to\infty$. 

Henceforth, we assume that Eqs.~(\ref{2b-1})-(\ref{cond}) are fulfilled. At this stage it is important to note that, by assuming Eqs.~(\ref{2b-1})-(\ref{2b-3}) for the variations of $a_n$, $k_n$ and $\omega_n$, we are not seeking to mimic any known wave spectrum. This is just a convenient way for us to test our theoretical predictions by varying a single parameter, $\gamma$. However, we note that the series $\sum_n (k_na_n)^2$ should converge to prevent the electrostatic energy from diverging with $N$, which requires $\gamma<1$. For this reason, in this article we restrict to $\gamma\leq 1$, and the case $\gamma=1$ is mainly considered to recall the results already obtained in Ref.~\onlinecite{benisti1}. 

We also note that Langmuir waves all have about the same frequency, close to the plasma frequency, while the wavenumbers are expected to be bounded from above. Indeed, it is very hard to excite Langmuir waves with large wavenumbers (compared to the inverse of the Debye length) due to the large Landau damping rate of such waves. Hence, for Langmuir waves, $\omega_n/k_n$ is expected to increase with $n$ and, for such waves, one may expect locality and quasilinear diffusion only if the series $\sum_n(k_na_n)^2$ converges. This condition should always be satisfied because it corresponds to the convergence of the electrostatic energy when $N\to\infty$. For ion acoustic waves, $\omega/k$ is a constant and the condition for locality and quasilinear diffusion is the convergence of the series $\sum_na_n^2$. However, once again, we do not focus here on any particular kind on plasma modes but rather seek to confirm the theoretical predictions of Section~\ref{theory} by varying one single parameter, $\gamma$.

As regards the numerical simulations, we use a leapfrog integrator~\cite{verlet} to solve the Hamilton equations for $H$, defined by Eq.~(\ref{H}). In all our simulations, we only consider the motion of one single particle with initial velocity, $v_0$, and compute averages with respect to the random phases $\varphi_n$. Moreover, although not shown here, we systematically check that we use a small enough time step, $\delta t$, for the simulation to have converged. Then, the number of phase realizations used for our averaging usually depends on the smallness of the time step, $\delta t$, we have to use. Moreover, we systematically stop the simulations before the orbits reach the phase space boundaries.

\subsection{Results when $\gamma=-1$}
\label{gamma-1}
According to Eq.~(\ref{gamma}), wave-particle interaction is not local when $\gamma=-1$ and the initial quasilinear regime is not expected. 
\subsubsection{$A=1$, $v_0=50$}
\label{A1}
\begin{figure}[!h]
\centerline{\includegraphics[width=12cm]{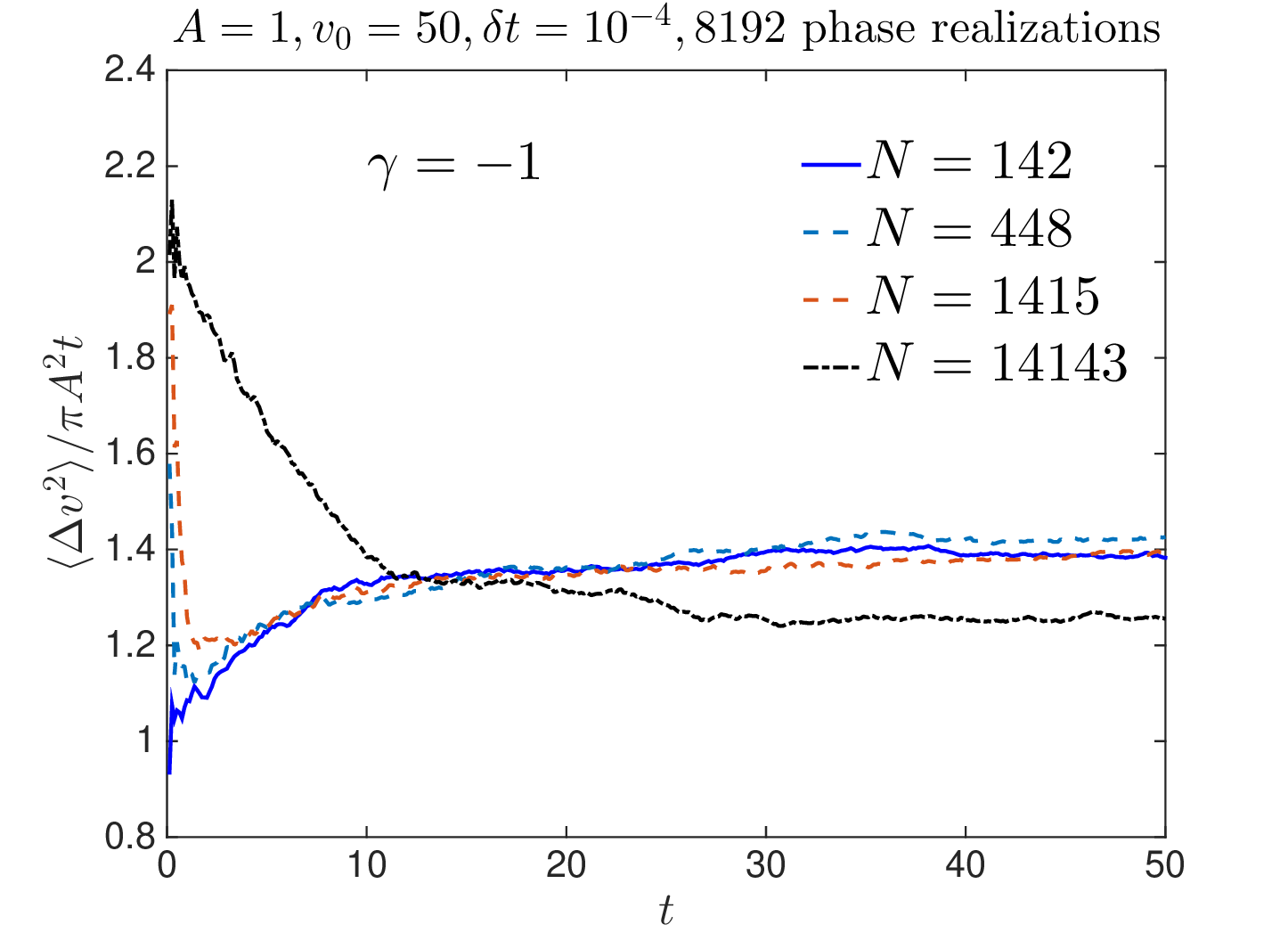}}
\caption{\label{f1}Time dependence of  $\langle\Delta v^2(t)\rangle/2  \DQL t=\langle\Delta v^2(t)\rangle/\pi A^2t$ when $\gamma=-1$, $A=1$ and $v_0=50$. The blue solid line is for $N=142$, the green dashed line for $N=448$, the red dashed line for $N=1415$ and the black dashed-dotted line for $N=14143$. {These values for $N$ have been chosen so that the maximum phase velocity be, respectively, 2, 20, 200 and $2\times10^4$ times $v_0$.} }
\end{figure}

As may be seen in Fig.~\ref{f1} when $A=1$ and $v_0=50$, the initial values of $\langle\Delta v^2(t)\rangle/2  \DQL t=\langle\Delta v^2(t)\rangle/\pi A^2t$ depend a lot on $N$, and get further away from unity (as would be expected for quasilinear diffusion) as $N$ increases. We did not find an initial regime of quasilinear diffusion except maybe when $N=142$, which is the smallest value for $N$ we investigated. Moreover, since the initial time variations of $\langle\Delta v^2(t)\rangle$ significantly depend on $N$, wave-particle interaction is clearly not local. 

\begin{figure}[!h]
\centerline{\includegraphics[width=12cm]{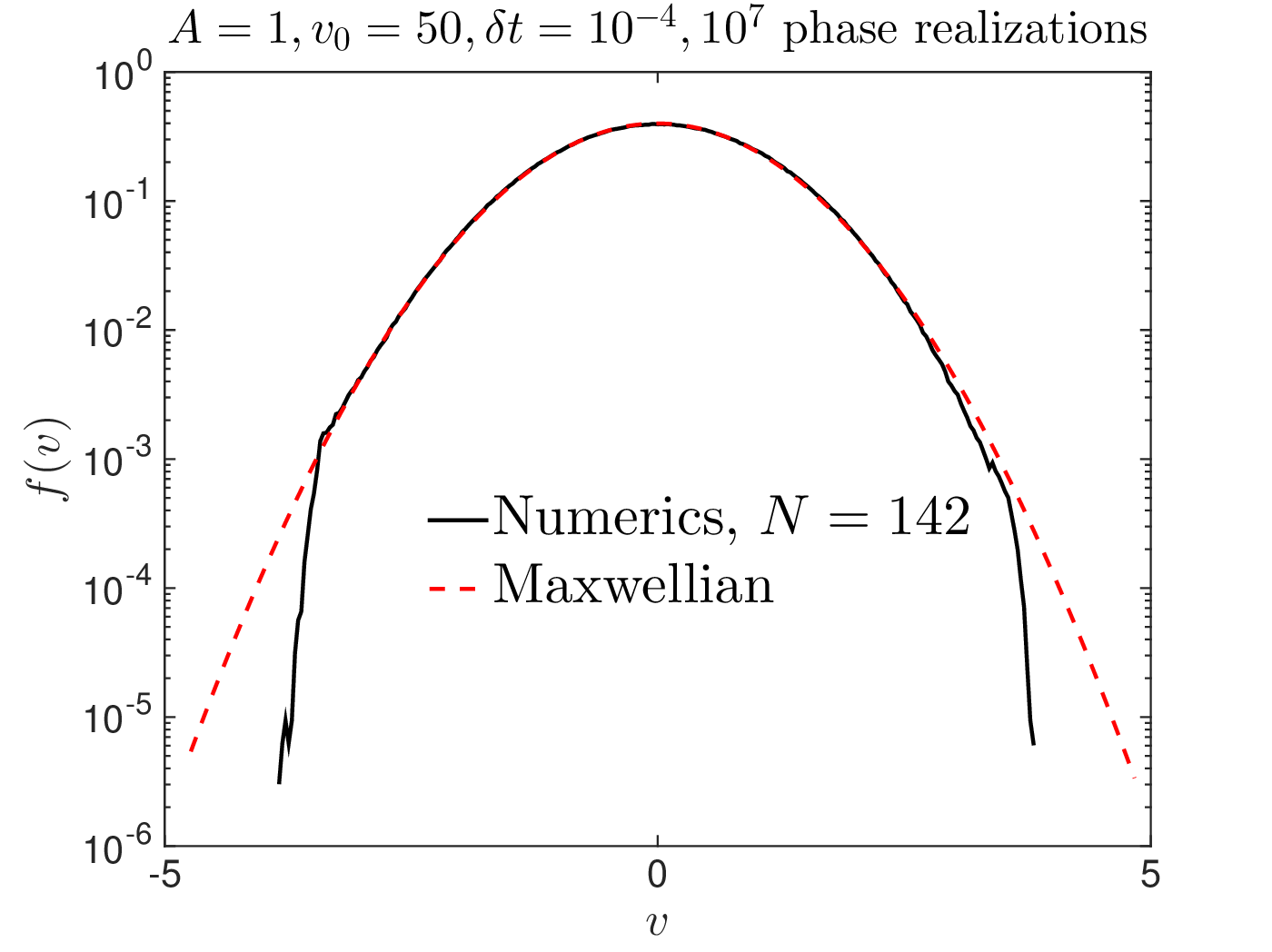}}
\caption{\label{f2}The black solid line shows the velocity distribution function, $f(v)$, found numerically when $\gamma=-1$, $v_0=50$, $N=142$, and at time $t=50$, by repeating the simulations with $10^7$ different phase realizations. $f(v)$ is normalized so that its mean is zero and its variance is unity. The red dashed line shows the Maxwellian with same mean and same variance. }
\end{figure}

Fig.~\ref{f1} also shows that $\langle\Delta v^2(t)\rangle/\pi A^2t$ eventually tends towards a constant, whatever the value $N$ we investigated. This seems to indicate an eventual chaotic diffusion, although wave-particle interaction is not local. This is confirmed by Fig.~\ref{f2} for $N=142$, which shows that the normalized velocity distribution is close to a Maxwellian. Moreover, when $N=142$, $N=448$ and $N=1415$, $\langle\Delta v^2(t)\rangle/\pi A^2t$ tends towards the same constant, close to 1.4. As may be seen in Fig.~\ref{f6}, this is the same result as that obtained when $\gamma=1$, which corresponds to the Hamiltonian $H_p$, defined by Eq.~(\ref{Ht}). However, when $N=14143$, $\langle\Delta v^2(t)\rangle/\pi A^2t$ tends towards a constant close to 1.2, which underlines the non locality of wave-particle interaction.
\subsubsection{$A=11.5$, $v_0=50$}
\label{A11.5}
\begin{figure}[!h]
\centerline{\includegraphics[width=12cm]{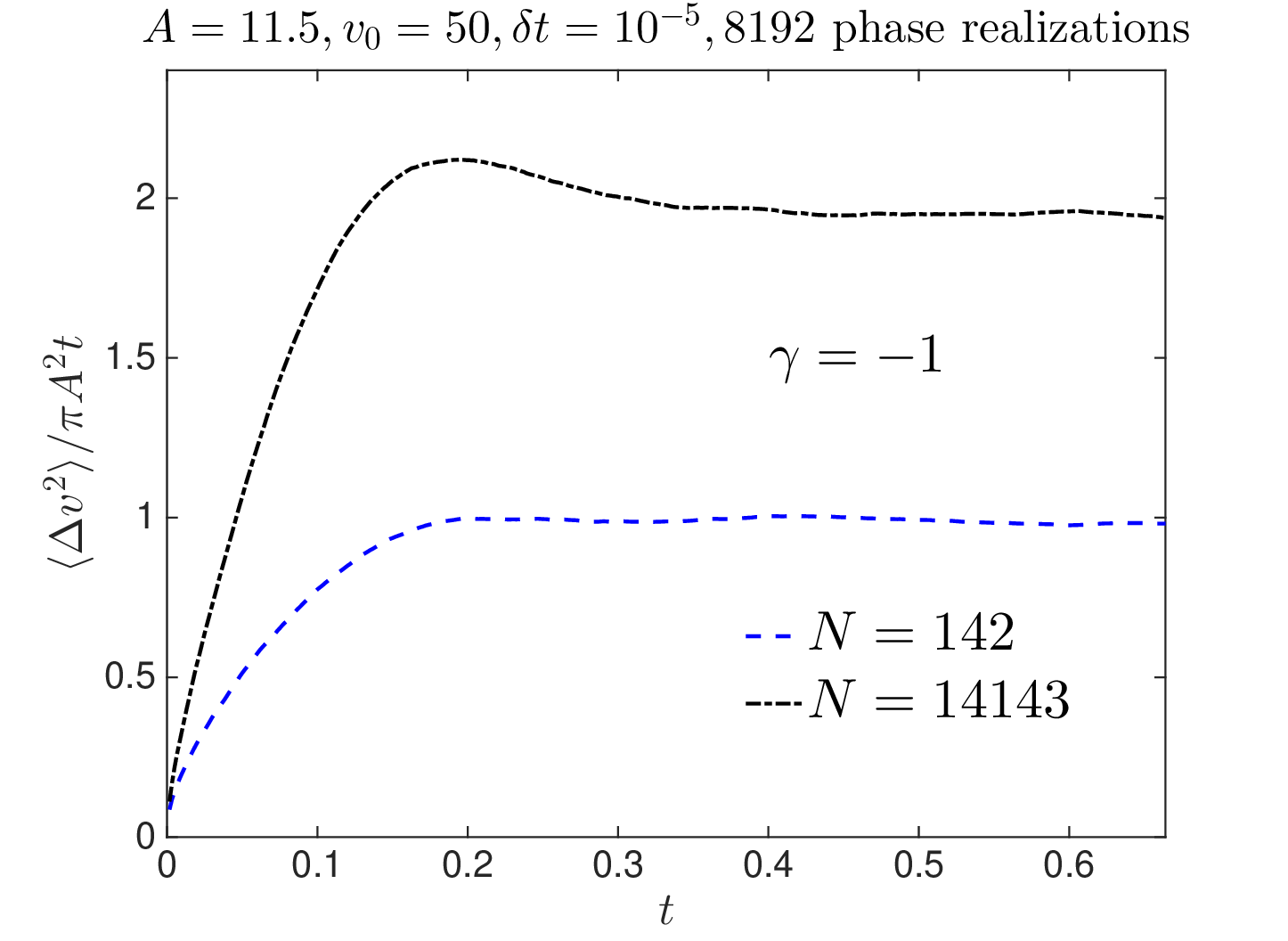}}
\caption{\label{f3}Time dependence of  $\langle\Delta v^2(t)\rangle/2  \DQL t=\langle\Delta v^2(t)\rangle/\pi A^2t$ when $\gamma=-1$, $A=11.5$ and $v_0=50$. The blue dashed line is for $N=142$ and the black dashed-dotted line for $N=14143$.}
\end{figure}

Fig.~\ref{f3} shows the time dependence of $\langle\Delta v^2(t)\rangle/\pi A^2t$ when $v_0=50$ and $A=11.5$. This value for $A$ is interesting because, as may be seen in Fig.~\ref{f7}, for the Hamiltonian $H_p$ corresponding to $\gamma=1$, $\langle\Delta v^2(t)\rangle$ keeps on evolving in a quasilinear way (until the particles eventually reach the phase space boundaries). In Ref.~\onlinecite{benisti1}, using locality, this result was interpreted as the crossover between the initial quasilinear regime of diffusion and the later, chaotic one. When wave-particle interaction is no longer local, this crossover is not expected to occur, and diffusion may be different from quasilinear. This may be appreciated in Fig.~\ref{f3} showing that, when $N=14143$, diffusion is at a rate twice faster than quasilinear. However, when $N=142$ diffusion seems to be, indeed, quasilinear.
\clearpage

\subsection{Results when $\gamma=-2$}
\label{gamma-2}
In order to show that the non-occurrence of the initial quasilinear regime and the non locality of wave-particle interaction are not specific to the situation corresponding to $\gamma=-1$, we now show similar results as those of Figs.~\ref{f1} and \ref{f3} but for $\gamma=-2$. 
\subsubsection{$A=1$, $v_0=50$}
\label{A1b}
\begin{figure}[!h]
\centerline{\includegraphics[width=12cm]{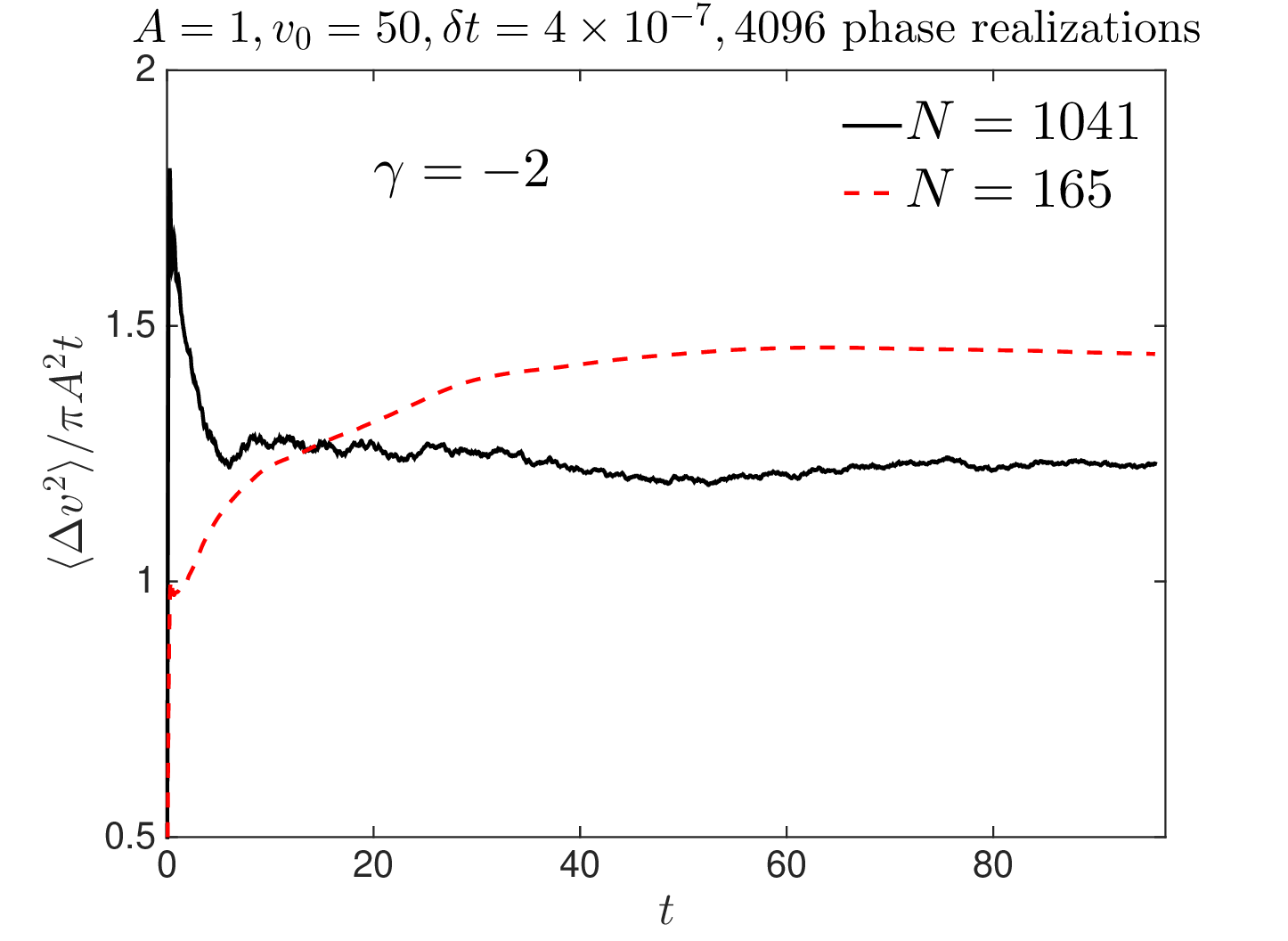}}
\caption{\label{f4}Time dependence of  $\langle\Delta v^2(t)\rangle/2  \DQL t=\langle\Delta v^2(t)\rangle/\pi A^2t$ when $\gamma=-2$, $A=1$ and $v_0=50$. The red dashed solid line is for $N=165$, the black solid line for $N=1041$. These values for $N$ correspond to a maximum phase velocity being, respectively, 2 and 200 times $v_0$.}
\end{figure}

Fig.~\ref{f4}~shows the time evolution of $\langle\Delta v^2(t)\rangle/\pi A^2t$ when $A=1$ and $v_0=50$. Hence, it is the counterpart of Fig.~\ref{f1} with $\gamma$ changed from -1 to -2. For the smaller value of $N$ plotted in Fig.~\ref{f4}, $N=165$,  there is an initial quasilinear regime of diffusion which, however, does not exist when $N=1041$. Moreover, at late time chaotic diffusion seems to occur whether $N=165$ or $N=1041$, but the diffusion coefficient is larger when $N=165$. Hence, this shows again that wave-particle interaction is not local and that there is not always an initial regime of quasilinear diffusion.

The results shown in Fig.~\ref{f4} are similar to those obtained when $\gamma=-1$ for the same value of $A$ and $v_0$. An initial quasilinear regime for small values of $N$ but not for larger ones, and a chaotic diffusion coefficient larger for smaller values of $N$. The initial quasilinear diffusion for small values of $N$ may be understood from Eq.~(\ref{II-16}). Indeed, if the Hamiltonian only includes a small number of waves with phase velocities close to the particle velocity, quasilinear diffusion is expected. However, with a larger number of waves leading to large values of $B_n$ in Eq.~(\ref{II-16}) for phase velocities away from the particle velocity, there would clearly not be any quasilinear diffusion. As regards the fact that the chaotic diffusion coefficient is smaller for larger values of $N$, we do not have any explanation for this result, all the less that for larger wave amplitudes the magnitude of the diffusion coefficient may increase with $N$, as illustrated in Fig.~\ref{f3}. 

\subsubsection{$A=11.5$, $v_0=200$}
\label{A11.5b}
\begin{figure}[!h]
\centerline{\includegraphics[width=12cm]{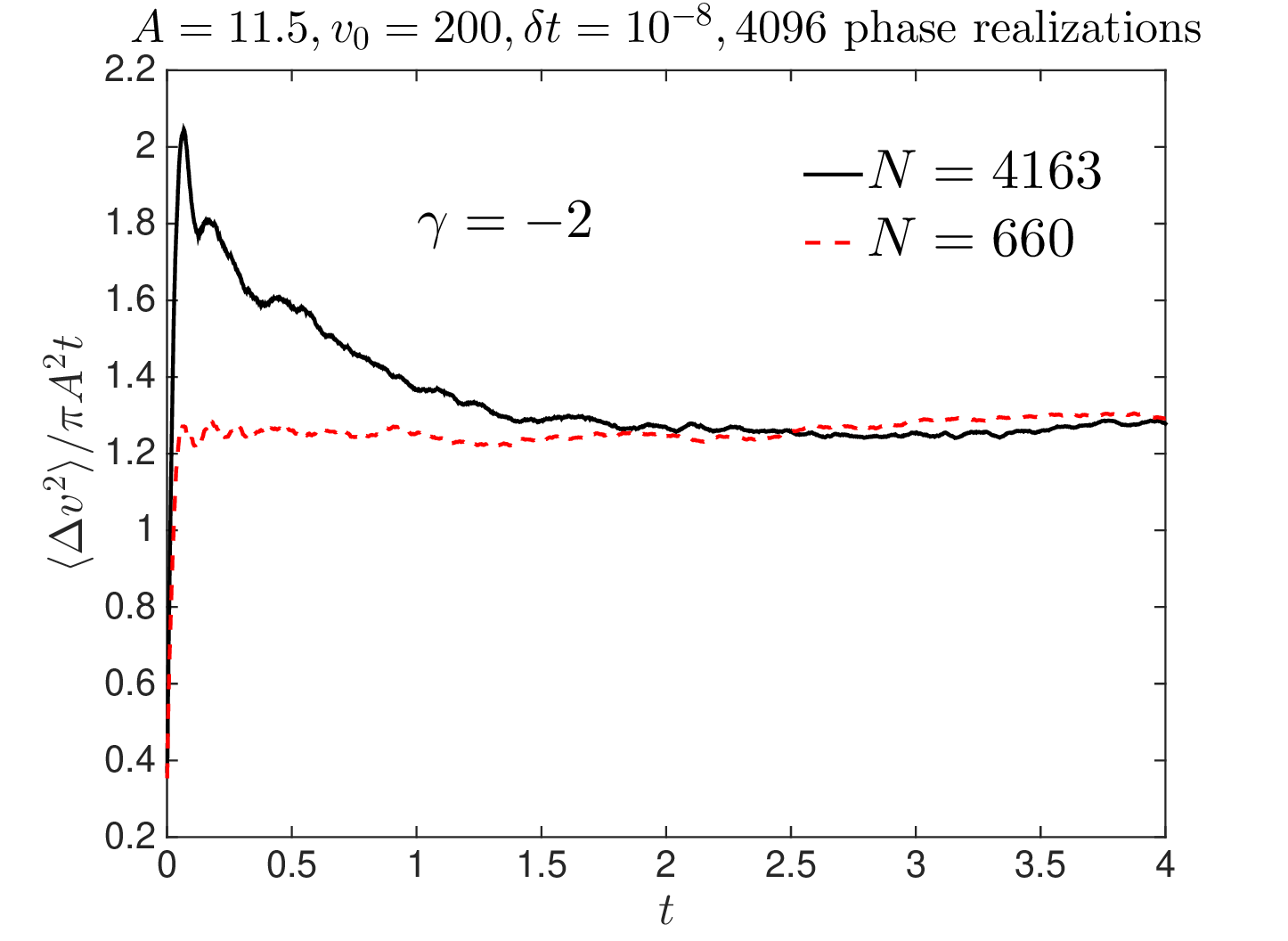}}
\caption{\label{f5}Time dependence of  $\langle\Delta v^2(t)\rangle/2  \DQL t=\langle\Delta v^2(t)\rangle/\pi A^2t$ when $\gamma=-2$, $A=11.5$ and $v_0=200$. The red dashed solid line is for $N=660$, the black solid line for $N=4163$. These values for $N$ correspond to a maximum phase velocity being, respectively, 2 and 200 times $v_0$.}
\end{figure}

Fig.~\ref{f5} shows the time dependence of $\langle\Delta v^2(t)\rangle/\pi A^2t$ when $v_0=200$ and $A=11.5$. For these parameters, we can never see any quasilinear diffusion, neither initially, nor at later times. Again, the early time dependence of $\langle\Delta v^2(t)\rangle$ is very sensitive to $N$. However, the final chaotic regime of diffusion is the same when $N=660$ and $N=4163$. Maybe it would change for larger values of $N$, as illustrated in Fig.~\ref{f1}. However, due to the very small time step we had to use, we did not try larger values of $N$. 

\subsection{Results when $\gamma=1$}
\label{gamma1}
The situation considered when $\gamma=1$ was studied in detail in Ref.~\onlinecite{benisti1} and the property of locality proved in Ref.~\onlinecite{jsp}. However, for the sake of completeness, we show some results for $\gamma=1$ which are the counterparts of those illustrated in Fig.~\ref{f1} and Figs.~\ref{f3} to \ref{f5}. 

\subsubsection{$A=1$, $v_0=50$}
\label{A1c}
\begin{figure}[!h]
\centerline{\includegraphics[width=12cm]{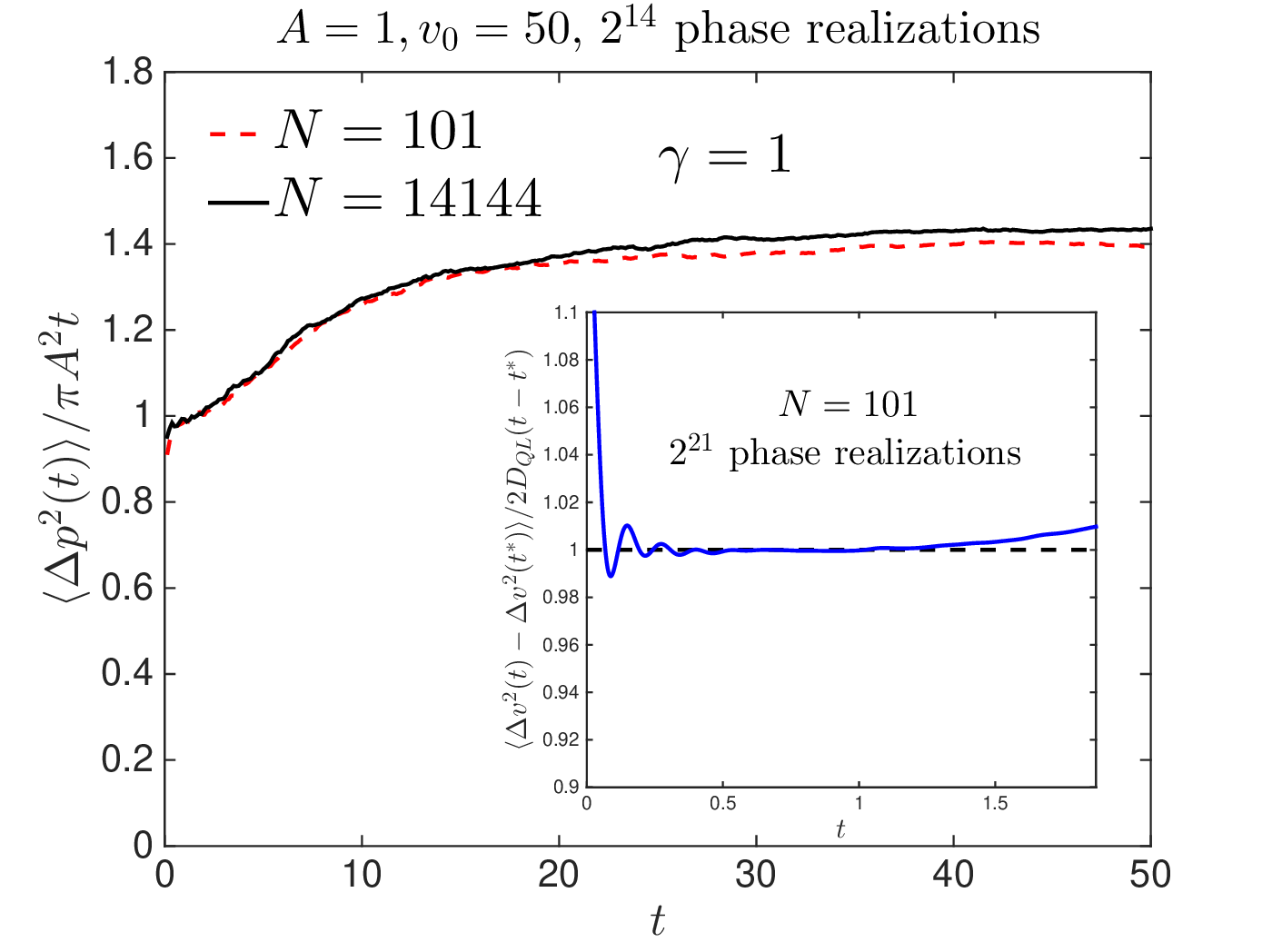}}
\caption{\label{f6}Time dependence of  $\langle\Delta v^2(t)\rangle/2  \DQL t=\langle\Delta v^2(t)\rangle/\pi A^2t$ when $\gamma=1$, $A=1$ and $v_0=50$. The red dashed solid line is for $N=101$, the black solid line for $N=14144$. The inset plots $\langle\Delta v^2(t)-\Delta v^2(t^*)\rangle/2  \DQL (t-t^*)$, where $t^*=2\pi/N$ and $N=101$, using $2^{21}$ phase realizations to compute the averaging. It clearly shows a quasilinear evolution (up to small oscillations) whenever $t^*\alt t\alt1.2$.}
\end{figure}

As may be clearly seen in Fig.~\ref{f6}, when $\gamma=1$, we do find an initial quasilinear regime when $A=1$ and $v_0=50$ and the time evolution of $\langle\Delta v^2(t)\rangle$ looks essentially independent of $N$, up to some fluctuations due to the finite number of phase realizations used in our averages. 

Actually, the early time evolution of $\langle\Delta v^2(t)\rangle$ in Fig.~\ref{f6} may be somewhat misleading since it seems to linearly increase with time. This is because the non-chaotic quasilinear evolution of $\langle\Delta v^2(t)\rangle$ only settles when $t>t^*$ where $t^*\approx2\pi/N$.  Indeed, only when $t\agt t^*$ may the sum of cosines in the right-hand side of Eq.~(\ref{II-16}) be approximated by a $2\pi$-periodic Dirac distribution. Then, when $t\agt t^*$, $\langle\Delta v^2(t)\rangle$ reads $\langle\Delta v^2(t)\rangle\approx2D_{QL}(t-t^*)+\langle\Delta v^2(t^*)\rangle$ and $\langle\Delta v^2(t^*)\rangle\neq 2D_{QL}t^*$ so that $\langle\Delta v^2(t)\rangle/2D_{QL}t$ may differ from unity for short times. In order to circumvent this difficulty and clearly evidence the early, non-chaotic, quasilinear diffusion, we plot in the inset of Fig.~\ref{f6}, $\langle\Delta v^2(t)-\Delta v^2(t^*)\rangle/2D_{QL}(t-t^*)$ when $N=101$, with $t^*\approx 2\pi/N$. This quantity is indeed close to unity when $t\agt t^*$ and for short enough times ($t\alt1.2$).

\subsubsection{$A=11.5$, $v_0=100$}
\label{A11.5c}
\begin{figure}[!h]
\centerline{\includegraphics[width=12cm]{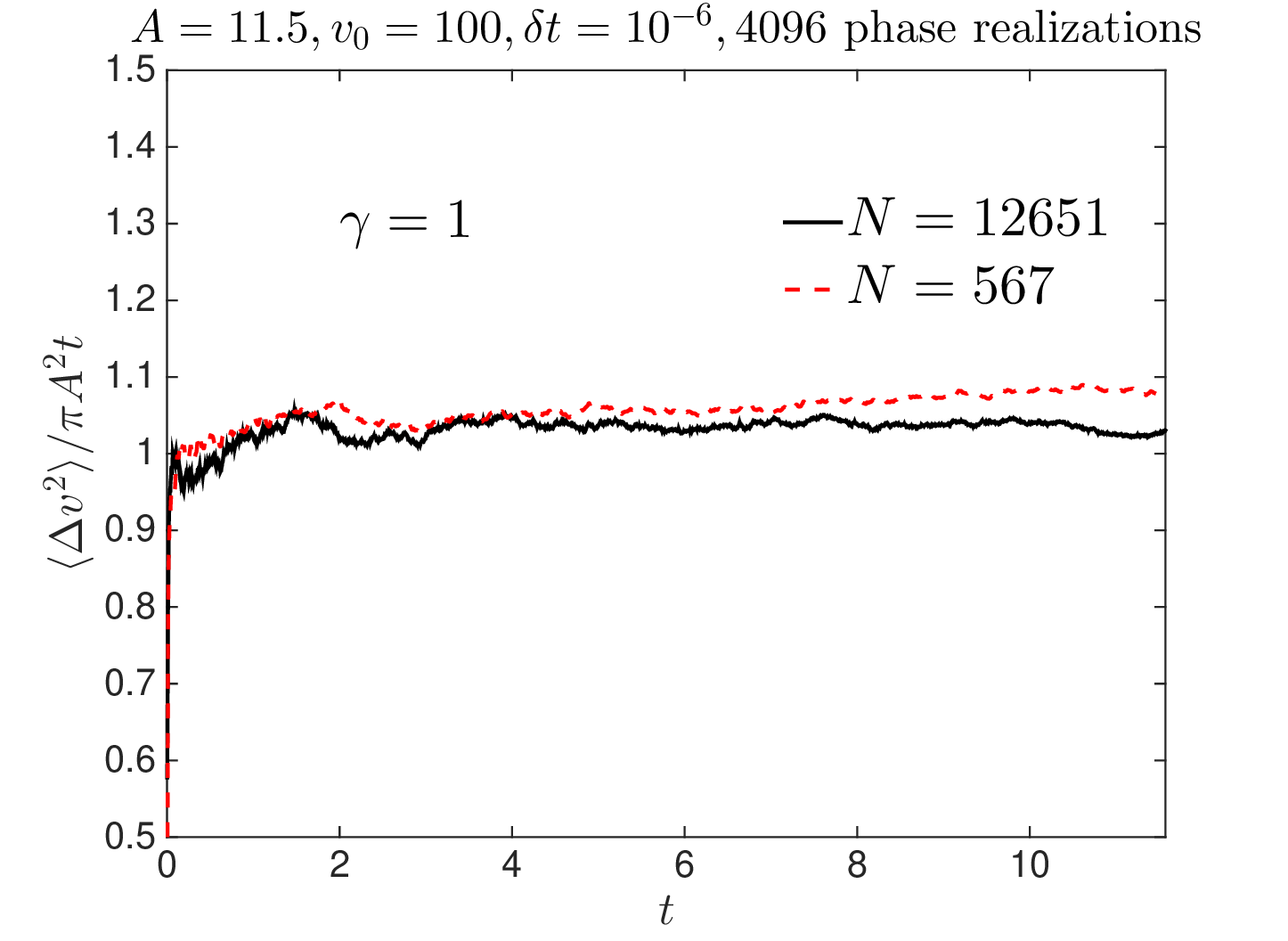}}
\caption{\label{f7}Time dependence of  $\langle\Delta v^2(t)\rangle/2  \DQL t=\langle\Delta v^2(t)\rangle/\pi A^2t$ when $\gamma=1$, $A=11.5$ and $v_0=100$. The red dashed solid line is for $N=567$, the black solid line for $N=12651$.}
\end{figure}

When $A=11.5$ and $v_0=100$, the time evolution of $\langle\Delta v^2(t)\rangle$ looks essentially independent of $N$ and corresponds to a quasilinear diffusion throughout the simulation, as shown in Fig.~\ref{f7}. 

As already mentioned in Paragraph~\ref{gamma-1}, this was interpreted in Ref.~\onlinecite{benisti1} as the direct consequence of locality leading to the crossover between the initial regime of quasilinear diffusion and 
the later regime of chaotic diffusion.
\clearpage

\subsection{Results when $\gamma=0.5$}
\label{gamma0.5}
The results corresponding to $\gamma=0.5$ are interesting for several reasons. They allow us to test our theoretical predictions that the property of locality and the initial quasilinear regime exist for values of $\gamma\neq 1$, so that there is nothing special about the Hamiltonian $H_p$ corresponding to $\gamma=1$. Moreover, the situation corresponding to $\gamma=0.5$ is somewhat more physical to that corresponding to $\gamma=1$ because $\sum(k_na_n)^2$ remains bounded when $N\to\infty$ so that the total electrostatic energy does not diverge with $N$. Furthermore, the results obtained with $\gamma=0.5$ allow us to check whether Eqs.~(\ref{2b-29})-(\ref{2b-32}) and Eq.~(\ref{w0}) do define universal transport properties when locality is effective, i.e., that the time evolution of $\dvt$ when $\gamma=0.5$ should be essentially the same as that for $\gamma=1$.  
\subsubsection{$A=1$, $v_0=50$}
\label{A1e}
\begin{figure}[!h]
\centerline{\includegraphics[width=12cm]{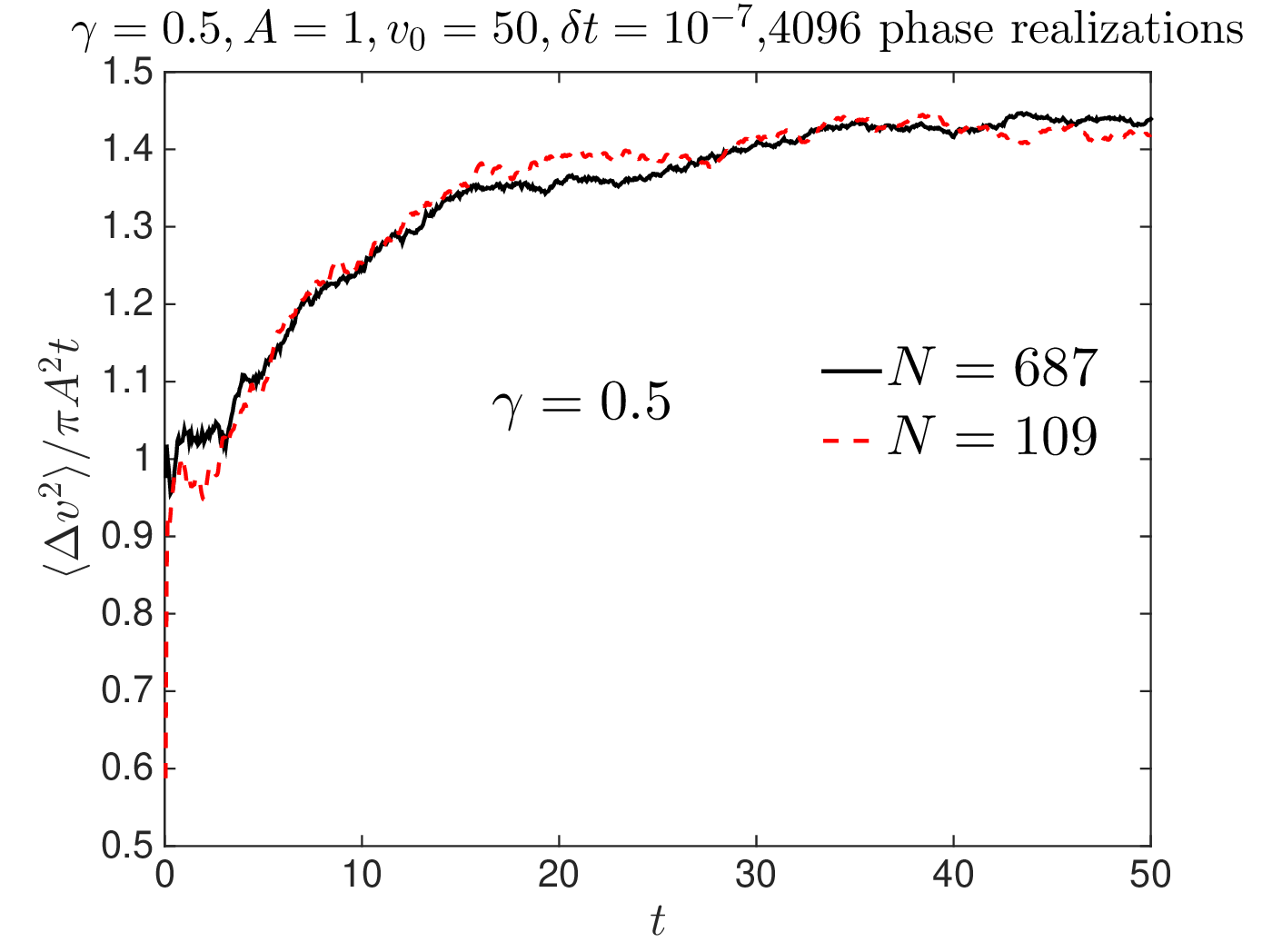}}
\caption{\label{f001}Time dependence of  $\langle\Delta v^2(t)\rangle/2  \DQL t=\langle\Delta v^2(t)\rangle/\pi A^2t$ when $\gamma=0.5$, $A=1$ and $v_0=50$. The red dashed solid line is for $N=109$, the black solid line for $N=687$. These values for $N$ correspond to a maximum phase velocity being, respectively, 2 and 20 times $v_0$.}
\end{figure}
Figure~\ref{f001} compares the time evolution of $\dvt/2  \DQL t$ when $A=1$ and $N=109$ with that when $N=687$. This figure clearly shows that, regardless of the value of $N$, there is an initial quasilinear regime of diffusion. Moreover, this figure also shows that the time variations of $\dvt$ are nearly the same when $N=109$ and $N=687$, thus evidencing that wave-particle interaction is local in velocity. 

\subsubsection{$A=11.5$, $v_0=50$}
\label{A11.5e}
\begin{figure}[!h]
\centerline{\includegraphics[width=12cm]{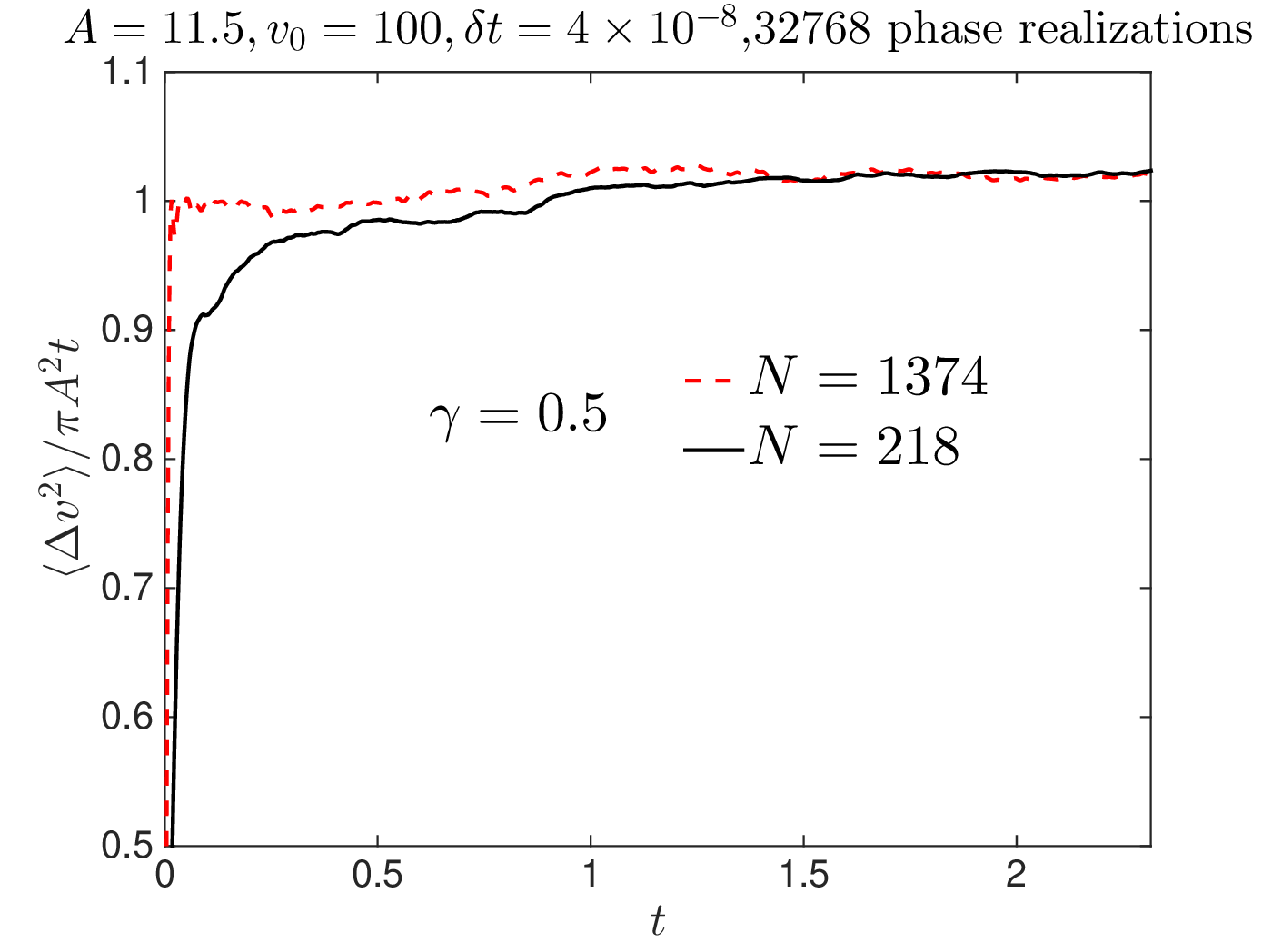}}
\caption{\label{f002}Time dependence of  $\langle\Delta v^2(t)\rangle/2  \DQL t=\langle\Delta v^2(t)\rangle/\pi A^2t$ when $\gamma=0.5$, $A=11.5$ and $v_0=100$. The red dashed solid line is for $N=218$, the black solid line for $N=1374$. These values for $N$ correspond to a maximum phase velocity being, respectively, 2 and 200 times $v_0$.}
\end{figure}
Figure~\ref{f002} compares the time evolution of $\dvt/2  \DQL t$ when $A=11.5$ and $N=218$ with that when $N=1374$. As may be seen in this figure, for these two values of $N$ the time variation of $\dvt$ is very close to that corresponding to a quasilinear diffusion. This is the same result as that found when $\gamma=1$, as expected.
\clearpage

\subsection{Comparisons between the transport properties corresponding to $\gamma=0.5$ and $\gamma=1$.}
\label{comparegamma}
\begin{figure}[!h]
\centerline{\includegraphics[width=15cm]{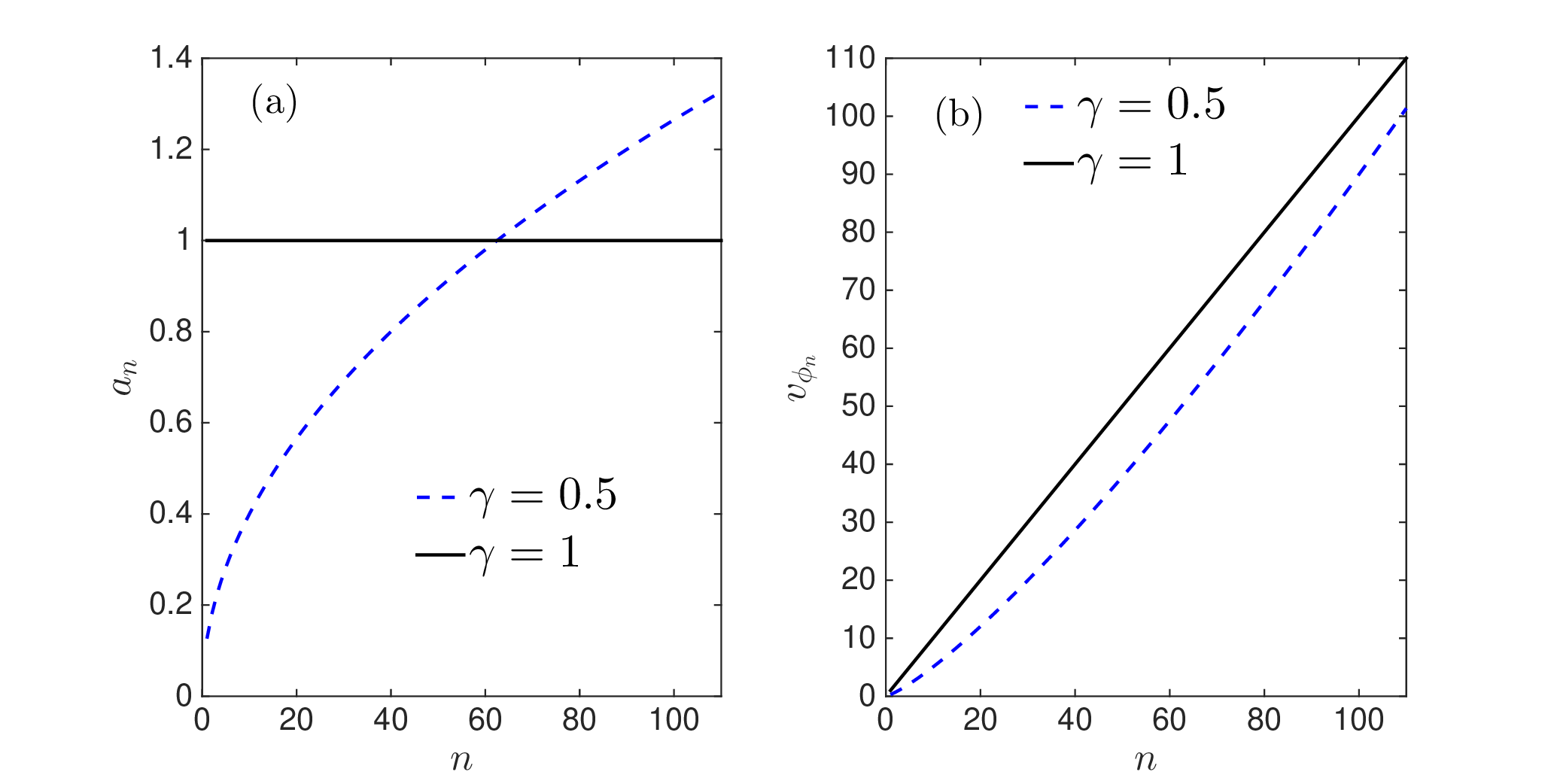}}
\caption{\label{anvn} Modes amplitudes [Panel (a)] and phase velocities [Panel (b)] when $\gamma=1$ (black solid lines) and $\gamma=0.5$ (blue dashed line). }
\end{figure}

In this Section, we compare the transport properties of two wave spectra corresponding to $\gamma=0.5$ and $\gamma=1$, which differ as regards the modes amplitudes and phase velocities, as may be appreciated in Fig.~\ref{anvn}. 

\begin{figure}[!h]
\centerline{\includegraphics[width=12cm]{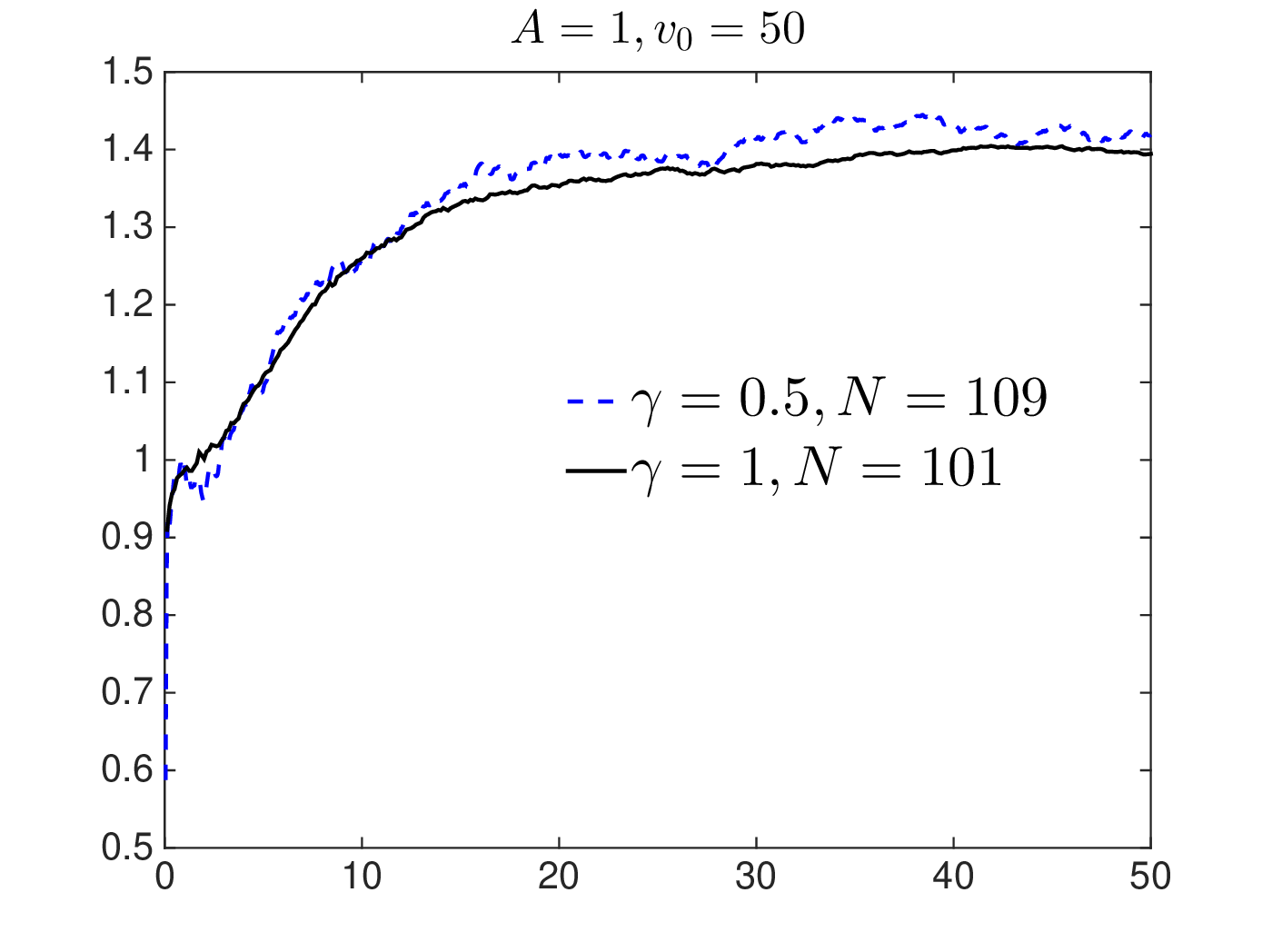}}
\caption{\label{f003}Time dependence of  $\langle\Delta v^2(t)\rangle/2  \DQL t=\langle\Delta v^2(t)\rangle/\pi A^2t$ when $A=1$ and $v_0=50$, and when $\gamma=0.5$ (blue dashed line) and $\gamma=1$ (black solid line). }
\end{figure}

\begin{figure}[!h]
\centerline{\includegraphics[width=12cm]{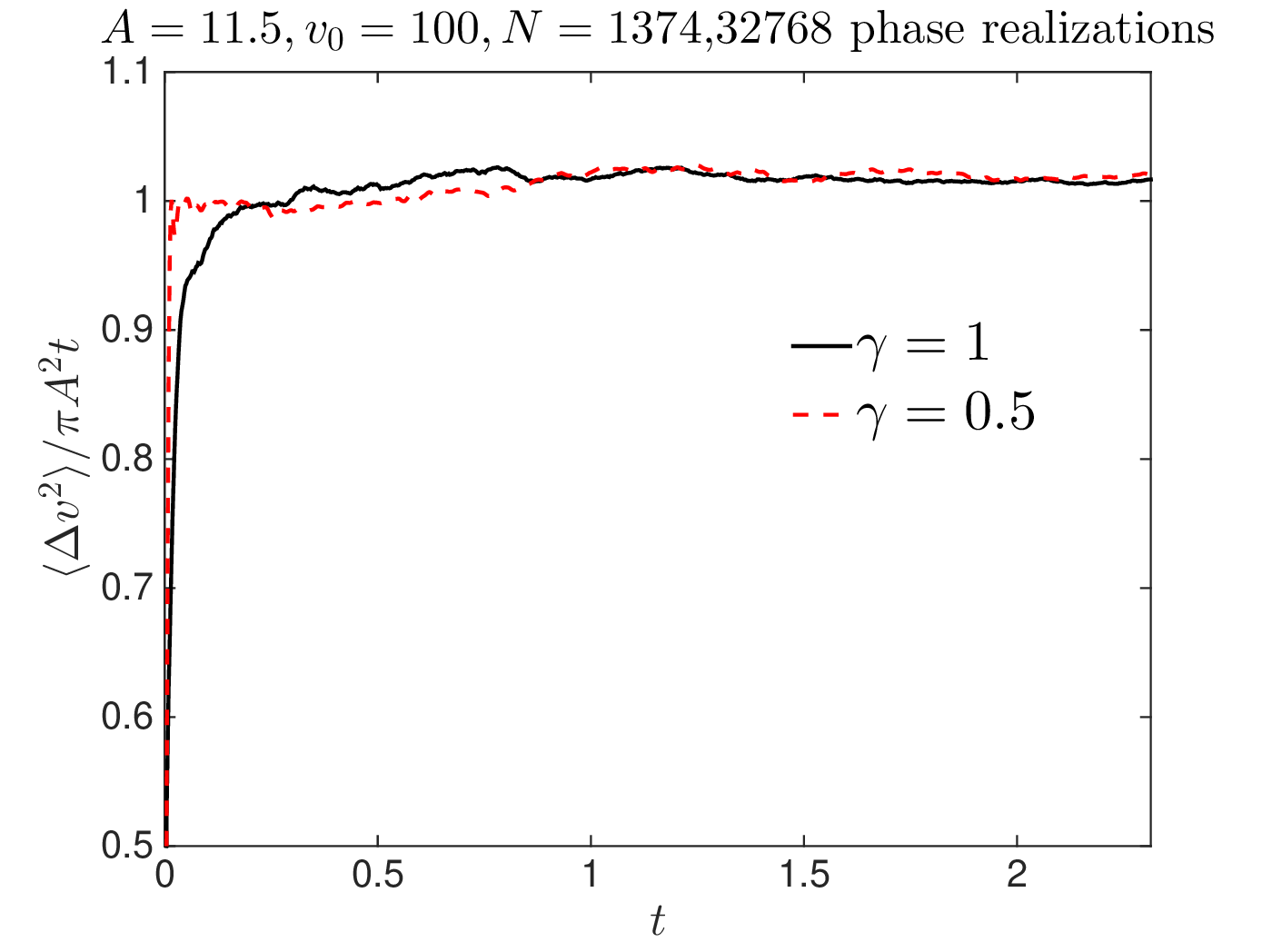}}
\caption{\label{f004}Time dependence of  $\langle\Delta v^2(t)\rangle/2  \DQL t=\langle\Delta v^2(t)\rangle/\pi A^2t$ when $A=11.5$, $v_0=100$,  $N=1374$ and when $\gamma=1$ (black solid  line) and  $\gamma=0.5$ (red dashed line). }
\end{figure}

In spite of these differences, Figs.~\ref{f003}~and~\ref{f004} respectively for $A=1$ and $A=11.5$ show that the time evolution of $\dvt$ is nearly the same for both spectra. This is a numerical evidence that, when wave-particle interaction is local, Eqs.~(\ref{2b-29})-(\ref{2b-32}) and Eq.~(\ref{w0}) do provide the conditions for universal transport properties. 
This also shows the relevance of the results obtained in Ref.~\onlinecite{benisti1} based only on the dynamics derived from Hamilton $H_p$. 
\clearpage

\section{Chaotic transport modeled as a diffusion process limited by phase space boundaries}
\label{transport}
In this section, we introduce a heuristic simple formula to derive the time evolution of the velocity distribution function in the situation when there is bounded chaotic transport, as defined in Section~\ref{poincare}. We restrict to regimes corresponding to Figs.~\ref{f7} and \ref{f002}, i.e., a quasilinear diffusion as long as  the orbits remain far enough from the phase space boundaries. However, we aim to derive the time evolution of the distribution function for arbitrarily long times. This is not a simple task since the exact equation we have to solve is actually unknown~\cite{note2}. In particular, although Dirichlet boundary conditions seem most appropriate for this problem, solving the diffusion equation with such boundary conditions would not yield the correct answer because $\int f(v)dv$ would not be conserved.  Hence, we only propose here a heuristic formula for the time evolution of the distribution function, that we compare against results from test particles simulations. Clearly, this is a preliminary study that would require a more in-depth investigation,  to better understand the range of validity of our formula 	and to address the more general situation when the diffusion coefficient is not uniform. However, we found useful to include our formula in this article due to its very good agreement with the numerical results. Hopefully, it will help improve the modeling of bounded chaotic transport. 

If the time evolution of the velocity distribution function, $f(v,t)$, resulted from a quasilinear diffusion, then
\begin{equation}
\label{sol2}
f(v,t)=\int_{-\infty}^{+\infty}f(v',0)\frac{\exp[-(v-v')^2/4\DQL t]}{\sqrt{4\pi\DQL t}}dv'.
\end{equation}
However, for a bounded chaotic transport, diffusion is only effective between the velocities $v_{\min}$ and $v_{\max}$ which are, respectively, the minimum and maximum velocities a particle may access. These are the minimum and maximum velocities of the domain illustrated in Fig.~\ref{p1}. Then, the number of particles within this domain must remain constant, i.e., $\int_{v_{\min}}^{v_{\max}}f(v,t)dv=\int_{v_{\min}}^{v_{\max}}f(v,0)dv$. This leads us to the following heuristic formula for $f(v,t)$,
\begin{equation}
\label{genial}
f(v,t)=\int_{v_{\min}}^{v_{\max}}f(v',0)\frac{\exp[-(v-v')^2/4\DQL t]}{\int_{v_{\min}}^{v_{\max}}\exp[-(y-v')^2/4\DQL t]dy}dv'.
\end{equation}
{Note that, when $\sqrt{2\DQL t}\ll (v_{\max}-v_{\min})$ and when $v_{\max}-v\gg\sqrt{2\DQL t}$ and $v-v_{\min}\gg\sqrt{2\DQL t}$, the integral in the denominator of the right-hand side of Eq.~(\ref{genial})  is close to $\sqrt{4\pi\DQL t}$ so that Eq.~(\ref{sol2}) is recovered, as should be. When $\sqrt{2\DQL t}\gg (v_{\max}-v_{\min})$, Eq.~(\ref{genial}) predicts a nearly constant $f(v)$ within the interval $v_{\min}\leq v\leq v_{\max}$, as expected and as may be appreciated in Fig.~\ref{fQL}.} As for $v_{\min}$ and $v_{\max}$ they are estimated the following way. Let $n_{\min}$ and $n_{\max}$ be such that  $\min(\omega_n/k_n)=\omega_{n_{\min}}/k_{n_{\min}}$ and $\max(\omega_n/k_n)=\omega_{n_{\max}}/k_{n_{\max}}$. Then, in Eq.~(\ref{genial}), we use the following expressions for $v_{\min}$ and $v_{\max}$,
\begin{eqnarray}
v_{\min}&=&\omega_{n_{\min}}/k_{n_{\min}}-2\sqrt{A_{n_{\min}}},\\
v_{\max}&=&\omega_{n_{\max}}/k_{n_{\min}}+2\sqrt{A_{n_{\max}}}.
\end{eqnarray}
\begin{figure}[h!]
\includegraphics[width=10cm]{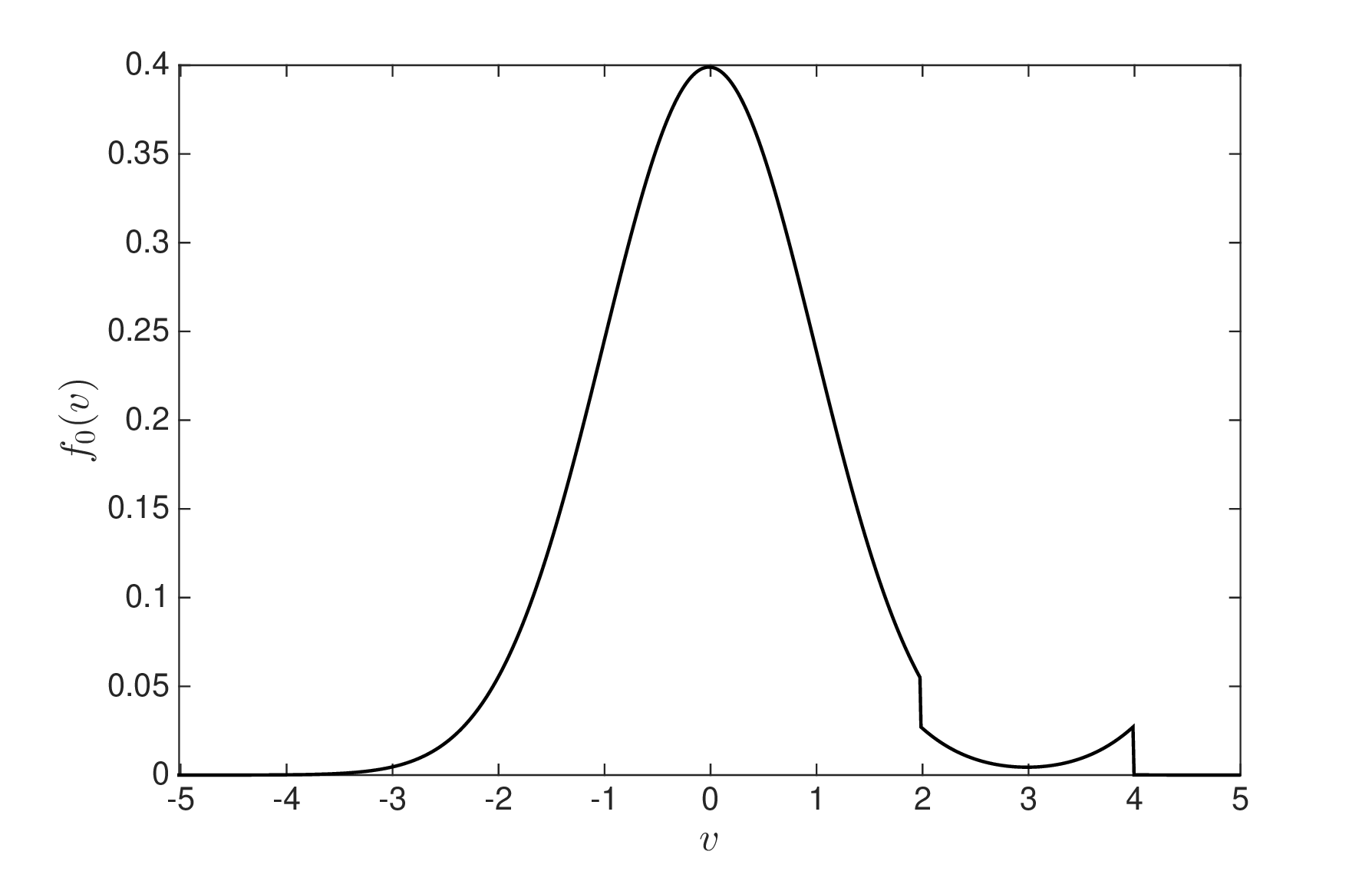}
\caption{\label{f0}  Initial velocity distribution function used to test the accuracy of Eq.~(\ref{genial}).}
\end{figure}

\begin{figure}
\centerline{\includegraphics[width=20cm]{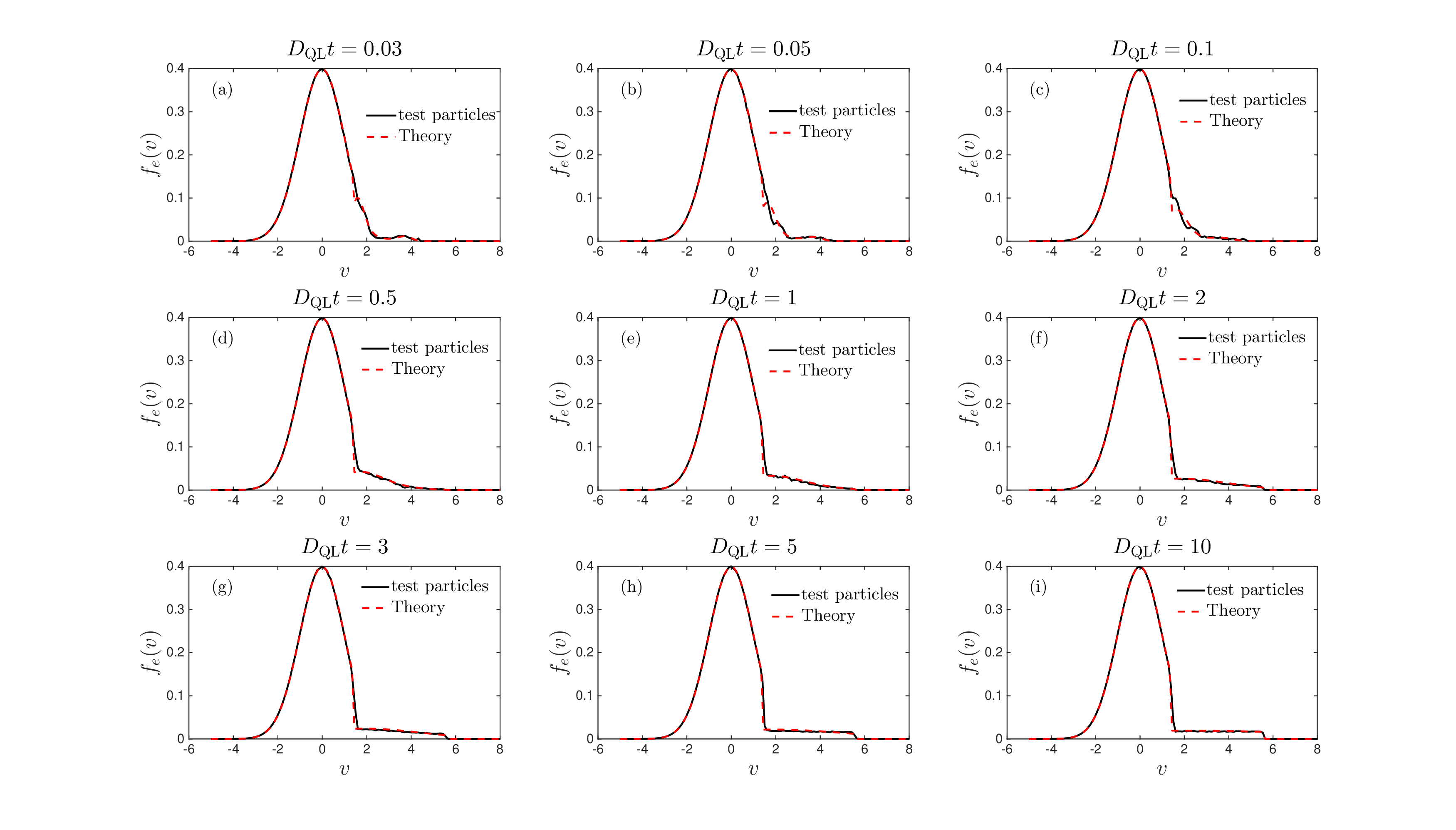}}
\caption{\label{fQL}  The red dashed curves represent the velocity distribution function as  derived from Eq.~(\ref{genial}) where $f_0(v)\equiv f(v,0)$ is illustrated in Fig.~\ref{f0}. The black solid lines plot the distribution function resulting from test particle simulations by solving the equations of motion derived from Hamiltonian $H_N$, Eq.~(\ref{Hnp}), for the same initial distribution function. }
\end{figure}

Fig.~\ref{fQL} compares results obtained from Eq.~(\ref{genial}) against those derived from test particles simulations. We use for $f(v,0)$ a Maxwellian,  $f(v,0)=e^{-v^2/2}/\sqrt{2\pi}$ when $v<2$ and $v>4$, and a symmetrized Maxwellian, $f(v,0)=\left[e^{-v^2/2}+e^{-(6-v)^2/2}\right]/2\sqrt{2\pi}$ when $\vert v-3\vert\leq1$. $f_0(v)\equiv f(v,0)$ is illustrated in Fig.~\ref{f0}. This is typically the distribution function resulting from the adiabatic trapping of electrons in a plasma wave~\cite{friou,benisti17}. Such a wave may be subjected to instabilities leading to the growth of sidebands~\cite{tacu}. Our work was motivated by the evolution of the electron distribution function under the action of these sidebands, whence the choice $f(v,0)$. As for the test particle simulations, we numerically solve the Hamilton equations of,
\begin{equation}
\label{Hnp}
H_N=\frac{v^2}{2}-\frac{11.5}{50^2}\sum_{n=80}^{270}\cos\left(x-\frac{n}{50}t+\varphi_n\right).
\end{equation}
This corresponds to a wave spectrum with phase velocities ranging from 1.6 to 5.4 and actually results from Hamiltonian $H_p$, Eq.~(\ref{Ht}), with $A=11.5$ by changing $t$ into $t/50$. For this value of $A$ we know that there is, indeed, quasilinear diffusion until the orbits reach the phase space boundaries. The change $t\to t/50$ induces a denser, and more physical, wave spectrum. Numerically, we calculate the motion of 524288 particles whose initial velocities are evenly distributed into 32768 values between -5 et 8 while their initial positions are evenly distributed into 16 values between $-7\pi/8$ et $\pi$. We use a leapfrog integrator~\cite{verlet} to calculate the electron motion with the timestep $\delta t=0.5$. The numerical distribution functions are actually histograms derived the following way. Each electron, $j$, with initial velocity, $v_j$, is assigned the weight $w_j=f_0(v_j)/16$, where $f_0(v)$ is the initial distribution function illustrated in Fig.~\ref{f0}. The velocity interval $[-5, 8]$ is divided into velocities $v_h^{(n)}$ separated by $\delta v_h^{(n)}\approx 0.1$. For each velocity $v_n=v_h^{(n)}+\delta v_h^{(n)}/2$ we estimate the distribution function the following way, $f_e(v_n,t)=\sum_jw_j$ where the sum is over all electrons $j$ with speed $v_j(t)$ such that $v_h^{(n)}\leq v_j(t)<v_h^{(n+1)}$. 

Fig.~\ref{fQL}~shows that Eq.~(\ref{genial}) provides results in very good agreement with those from test particle simulations except close to $v=2$ when $\DQL t\alt 0.5$. These slight discrepancies are expected because $f_0(v)$ is discontinuous at $v=2$ and a diffusion equation is not valid about a discontinuity for short times. This is easily proved the following way. Let us a consider a distribution function, $f_0(v)$, discontinuous at $v=v_d$, and let $\Delta f_0=f_0(v_d^-)-f_0(v_d^+)$ be the jump in $f_0$ at the discontinuity. Cleary, for any particle and at any time, the change in velocity induced by the wave spectrum remains finite. Consequently, the particles which have crossed the discontinuity by time $t$ have initial velocities, $v_0$, such that $\vert v_0-v_d\vert\leq\dvm(t)$, where $\dvm(t)$ is a finite velocity increment that grows with time and such that $\dvm(t=0)=0$. Then, at the discontinuity and at time $t$, the change in $f_0$ induced by the wave spectrum is
\begin{equation}
\label{dft}
\delta f(t)=\int_{v_d-\dvm(t)}^{v_d+\dvm(t)} f_0(v_0)\eta(v_0)dv_0,
\end{equation}
where $\eta(v_0)$ is the fraction of particles with initial velocity, $v_0$, which have crossed the discontinuity. Clearly, $\delta f(0)=0$ and, although $\delta f(t)$ is not necessarily a monotonic function of time, it would clearly tend to increase with $t$. Consequently, it requires a finite time for $\delta f$ to be larger than the initial jump in $f_0$, $\Delta f_0$. Before this time, the distribution function remains discontinuous. This is in contrast with the solution of a diffusion equation which becomes differentiable whenever $t>0$. 

In spite of the slight early discrepancies about the discontinuity, Fig.~\ref{fQL}~shows a very good agreement between the numerical and theoretical distribution functions whenever $\DQL t>0.5$, and the agreement becomes excellent whenever $\DQL t>3$. In particular, the transitions from a symmetric distribution to a nearly linearly decreasing one, and the later convergence towards a plateau, are very well reproduced by Eq.~(\ref{genial}) with the correct timescales. Hence, this equation may be used to describe bounded chaotic transport, at least when it may be modeled by a Hamiltonian as simple as $H_N$, whose universality has been discussed in Paragraph~\ref{comparegamma}. As already mentioned in the introduction of this Section, the generalization of Eq.~(\ref{genial}) to the situation when one cannot rely on a diffusion equation to model transport, and has to resort to a Fokker-Planck equation, is left for future work.
\section{Conclusion}

\label{conclusion}
In conclusion, this paper has addressed several important issues regarding phase space transport, in particular the relevance of its modeling as a quasilinear diffusion. As noted in the Introduction, quasilinear theory has been extensively discussed since its first publication in 1962~\cite{QL}. Nevertheless, to the best of the author's knowledge, these discussions had a limited impact on the plasma physics community. Quasilinear diffusion is still widely used to model transport because it is effective, rather simple, and usually gives the right orders of magnitudes for the transport rates. Yet, it is not always valid.

This paper has provided a simple criterion for assessing its validity, which is a central issue in many areas of physics and, most notably, in plasma physics. For a smooth discrete wave spectrum, the perturbative regime of quasilinear diffusion, akin to that introduced in the original 1962 papers, is valid only when the series on the right-hand side of Eq.~(\ref{eq16}) converges. This condition is equivalent to the requirement of locality in phase velocity for the wave-particle interaction. When this criterion is satisfied then, as illustrated in Figs.~\ref{f003} and \ref{f004} of Paragraph~\ref{comparegamma}, one can define a broad class of Hamiltonians whose transport properties are very similar to those of Hamiltonian $H_p$ [Eq.(\ref{Ht})]. When the condition for locality is not fulfilled, the early time evolution of the velocity spreading is not diffusive, as clearly shown in Fig.~\ref{f1} of Paragraph~\ref{A1}. Moreover, it depends on the number of modes in the discrete spectrum, unlike what quasilinear theory would predict. In spite of these very clear defects we have to acknowledge that, in all our simulations, quasilinear theory does provide the right order of magnitude for $\langle \Delta v^2(t)\rangle/t$ at short times, with discrepancies never exceeding a factor of about two.

Moreover, for longer times and when transport is effective, we numerically find that chaotic diffusion eventually seems to settle, even when wave-particle interaction is not local, as shown for example in Fig.~\ref{f2} of Paragraph~\ref{A1}. In all our simulations, the diffusion coefficient is always of the same order as the quasilinear one, again differing by no more than a factor of about two. However, when wave-particle interaction is not local, it depends on the number of modes. 

Now, as discussed in detail in Section~\ref{poincare}, diffusion cannot persist indefinitely  because phase space is bounded by KAM tori. In this paper, we accounted for these boundaries to derive an effective equation for the time evolution of the velocity distribution function, namely Eq.~(\ref{genial}) of Section~\ref{transport}. The predictions of this equation were compared against results from test particle simulations for an initially discontinuous distribution function. A very good agreement was found between the theoretical and numerical predictions, except in the vicinity of the discontinuity for short times. This was expected because, as proved in Section~\ref{transport}, a diffusion equation cannot accurately describe the early time evolution of a distribution function near its discontinuity. Given the high accuracy of the heuristic equation~(\ref{genial}), it would be of interest to further investigate its range of validity and to extend it to a non-uniform diffusion coefficient. This is left for future work. 

In summary, this paper has investigated transport induced by a discrete wave spectrum and introduced a simple, practical criterion for assessing the validity of quasilinear theory. Yet, it is difficult to discuss the notion of ``validity'' in very general terms. Indeed, as noted above, even when the velocity spreading is not diffusive, quasilinear theory appears to give the correct orders of magnitude for transport rates. This leads to a dual conclusion. On the one hand, quasilinear diffusion is not always valid and we showed that, for some wave spectra, the early time evolution of the distribution function is not even diffusive. On the other hand, if the goal is merely to estimate $\langle \Delta v^2(t) \rangle$, our numerical simulations suggest that quasilinear theory consistently provides the correct order of magnitude.

\begin{acknowledgments}
{The author acknowledges the very careful reading and comments by D. Escande and the useful discussions with Y. Elskens.}
\end{acknowledgments}

\section*{Author declarations}
\subsection*{Conflict of Interest}
The authors have no conflicts to disclose.
\section*{Data Availability}
The data that support the findings of this study are available from the corresponding author upon reasonable request.


\begin{thebibliography}{10}
\bibitem{note}
As noted in the derivation of Eq.~(\ref{new1}), provided that $B_l$ remains nearly constant whenever $-L\leq l\leq L$, with $L\gg 1$, the approximation $\sum_{l}B_l^2\cos[l(t'_1-t'_2)]\approx 2\pi\overline{B^2}\delta_{2\pi}(t_1-t_2)$ is only valid when $t\agt2\pi/L$. Then, the diffusive evolution of $\dvt$ only hods after this time interval. In most of this paper, we implicitly let this interval got to zero.
\bibitem{note2}
 One may actually think of modeling a bounded chaotic transport by a Fokker-Planck equation with $D$ nearly uniform when $v_{\min}\alt v\alt v_{\max}$ and $D=0$ whenever $v>v_{\max}$ or $v<v_{\min}$. We tried $D/\DQL=\min\left[(v-v_{\min})/v_0,1,(v_{\max}-v)/v_0Ê\right]$ with $v_0=0.1$ when $v_{\min}\leq v\leq v_{\max}$ and $D=0$ elsewhere. The distribution functions solving the Fokker-Planck equation (not shown here) are very similar to those derived from Eq.~(\ref{genial}), at least for the parameters of Fig.~\ref{fQL}. In particular, the same discrepancies with the results from test particle simulations are found for short times and the same very good agreement is recovered when $\DQL t\agt0.5$. However, Eq.~(\ref{genial}) provides an explicit analytical expression for $f(v,t)$ much easier to implement than the resolution of a Fokker-Planck equation.
\end{thebibliography}
\end{document}